\documentstyle[aps,prb]{revtex}
\begin{document}
\author{Roland Stumpf\,\cite{Albu} and Matthias Scheffler}
\address{Fritz-Haber-Institut der
  Max-Planck-Gesellschaft,\\ Faradayweg 4-6, D-14195 Berlin-Dahlem,
  Germany}
\title{
Ab-initio Calculations of Energies and Self-Diffusion on Flat and
Stepped Surfaces of Al and  their Implications on  Crystal Growth}

\date{\today}

\maketitle
\begin{abstract}
  Using density-functional theory we investigate several
properties of Al\,(111), Al\,(100), Al\,(110), and stepped
Al\,(111) surfaces.  We report results of formation energies
of surfaces, steps, adatoms, and vacancies.  For the
adsorption and diffusion of Al on flat regions of Al\,(111)
surfaces we find the hcp site energetically slightly preferred
over the fcc site. The energy barrier for
self-diffusion on Al(111) is very low (0.04\,eV).  Coming close to one of
the two sorts of close packed, monoatomic steps on Al\,(111),
labeled according to their \{111\} and \{100\} micro-facets,
Al adatoms experience an attraction of $\lesssim$~0.1\,eV
already before direct contact with the edge of the step. This
attraction has a range of several atomic spacings and is of
electronic origin.  Upon arrival at the lower step edge, the
adatom attaches with no barrier at a low energy five-fold
coordinated site.  Coming from the upper terrace, it
incorporates into the step by an atomic exchange process,
which has a barrier below $0.1\,$eV for both sorts of close
packed steps.  The barrier for diffusion along the lower edge
is 0.32\,eV at the \{100\}-faceted step and 0.39\,eV at the
\{111\}-faceted step.  Unexpectedly the latter diffusion process
proceeds by an exchange mechanism. Diffusion by a very similar exchange
mechanism is also found for the ``easy'' direction on the Al\,(110)
surface, i.\,e., along the channels. We show that Al\,(110) is a model
system for diffusion at the \{111\}-faceted step on Al\,(111) because of
its very similar local geometry.  Our results enable the estimate of
temperature ranges for different modes of homoepitaxial growth on
Al\,(111).  Of particular importance for the growth modes and the
resulting surface morphology are the rather low barriers for diffusion
across the descending steps and the rather high barriers for diffusion
along the steps.  We discuss the shapes of islands on Al\,(111) during
growth and in thermodynamic equilibrium. During growth (depending on the
temperature) the shape can be fractal, triangular, or hexagonal and mainly
determined by kinetics; in thermodynamic equilibrium the island shape is
hexagonal and determined by the different step formation energies.  Many
of the phenomena, which we predict for Al, were found for other metals
in experiment.
\end{abstract}

\pacs{68.35.Fx, 68.55.-a, 68.35.Md, 71.45.Nt}
\thispagestyle{empty}

\section{Introduction}

Atomic scale imaging techniques have proven to be useful tools to
improve the understanding of atomic processes at surfaces.  Prominent
examples for studies at metals are the field ion microscopy (FIM)
studies of self diffusion at transition metal
surfaces\cite{Ayrault74,Wang89,Kellogg91,Tsong91,Tsong90}, the electron
microscope studies of step structures and growth on halide,
semiconductor, and metal surfaces\cite{Bethge90,Klaua85}, and the
scanning tunneling microscopy (STM) studies of the epitaxial growth and
sputter removal of Pt(111).\cite{Kunkel90,Bott92,Michely93} Several
observations made in those studies like the reentrant layer-by-layer
growth at low temperatures or the temperature variation of the growth
form of islands at higher temperatures are not fully understood yet. For
other growth phenomena like the fractal island shape at low
temperatures\cite{Bott92,Hwang91} a satisfying explanation does exist.
Understanding the origins for different forms of growth bears
technological relevance. Typically it is desirable to have
layer-by-layer growth in order to produce high quality films and to
achieve this at not too high
temperatures.\cite{Egelhoff89,Hinch93,Oppo93}

How a surface develops during growth is a consequence of the microscopic
adatom-surface interaction, especially at binding sites and at
transition states of surface diffusion.\cite{Vineyard57,Gomer90} If the
rates for all relevant diffusion processes are known, the evolution of
the surface during growth can be
calculated.\cite{Venables84,hydro,Kenny92,Smilauer93} Because of the
computational effort required for a quantum-mechanical description of
the microscopic interaction, several quasi-classical methods have been
used in the past.\cite{Hammonds92,Liu91,Liu92,Nelson93,Hansen91}
However the reliability of these calculations have always been
questionable, especially because neither the influence of the kinetic
energy operator for the electrons nor self-consistent rearrangements of
the electron density were taken into account properly. The kinetic
energy of the electrons largely determines the nature of the chemical
bond by splitting the electronic energies into bonding and antibonding
levels, or by influencing the charge distribution at metal surfaces,
which is governed by the spill out of density into the vacuum and by the
reduction of the charge density corrugation
(Smoluchowski smoothing~\cite{Smoluchowski41,Zangwill88}).  All
quantum-mechanical effects which are considered to be relevant for
chemisorption are taken into account in density-functional
theory (DFT) to a high level
of accuracy when used together with the local-density
approximation (LDA) of the exchange-correlation functional.~\cite{Pickett89}

In this paper we report a rather extensive set of DFT-LDA calculations
of adsorption and diffusion of Al adatoms on different surfaces of fcc
aluminum.\cite{Scheffler92} We even include comprehensively the
diffusion at steps in our study of Al\,(111). The role of steps in
determining the growth morphology is known since long
time.\cite{Bethge90} We chose Al because it is a prototype of a
simple $s$-$p$ metal, hoping that the interpretation of
any observation would be particular clear and provides
insights  which are transferable to other systems.

Besides the flat (111), (100), and (110) surfaces, also the
two different close-packed steps on Al\,(111) are considered.
These steps are called $\langle 110\rangle/\{100\}$ and
$\langle 110\rangle/\{111\}$ according to the step's
orientation, which is the $\langle 110\rangle$ direction, and
the steepest micro-facet at their edges (see
Fig.~\ref{StepScheme}).\cite{Lang72,Michely91} The influence
of steps is of paramount importance for the description of
growth processes.\cite{Bott92,Smilauer93,Bethge90,Chen93} In
particular we like to understand the experimentally
established differences between these two sorts of steps on
(111) surfaces of fcc metals.  Their slightly different
geometry leads to different formation energies,\cite{Bott92}
to different diffusion mechanisms and energy
barriers,\cite{Michely93,Wang90,Wang91} and they are found to
have different dipole moments.\cite{Besocke77}

Using the calculated diffusion barriers and estimated diffusion
prefactors we estimated the temperature ranges for different growth
modes on Al\,(111).  To get a detailed description of the temperature
dependence of the growth of the Al\,(111) surface one should use our
results on surface diffusion as input for a
theory\cite{Venables84,hydro,Kenny92,Smilauer93} which solves for the
rate equations which determine the evolving surface morphology during
growth. This still remains to be done.

The paper is organized as follows. First we give a short description of
our {\em ab-initio\/} method and describe some technical aspects which
make it particularly efficient for the calculation of large metallic
systems.  Section~\ref{Differ} describes differences in the formation
energy of the two sorts of close-packed steps on Al\,(111).  In
Section~\ref{IndDip} we discuss the adatom- and step-induced dipole
moment on Al\,(111) and, connected to that, the
work function differences between Al\,(111), Al\,(100), and
Al\,(110).
In Section~\ref{Al111} the surface self-diffusion is
investigated, first on the flat Al\,(111) surface, then approaching a
step, and finally at the step. Also vacancy diffusion on the flat
Al\,(111) surface is considered. We compare the self-diffusion on
Al\,(110) with that on the stepped Al\,(111) surface and we compare
diffusion at the two different steps on Al\,(111). In Section~\ref{Unif}
we describe some uniformity in surface self-diffusion on fcc metals,
which help us to estimate diffusion prefactors for the self-diffusion on
Al surfaces.  Using the calculated diffusion barriers and the estimated
prefactors, we summarize in Section~\ref{Atom} our understanding of
the temperature dependence of atomic transport processes and of
homoepitaxial growth on Al\,(111).  In the appendix we also present some
results for the self-diffusion on Al\,(100).

\section{Theory}
\label{Theo}

The computer code, {\sf fhi93cp}, used in this study is described in
Ref.~\onlinecite{Stumpf94b}. We therefore summarize here only some
essentials of the method. We use density-functional theory and
treat the exchange-correlation functional in the local-density
approximation.\cite{Ceperley80} The Kohn-Sham
equations\cite{Kohn65} are solved by a Car-Parrinello-like iterative
scheme~\cite{Car85} using the steepest descent
approach~\cite{Williams87} for the update of the wavefunctions. We use a
fully separable {\em ab-initio\/} pseudopotential~\cite{Stumpf90} for Al
where the $d$-potential is treated as local and $s$ and $p$ potentials
are described by projection operators.  The electronic wavefunctions are
expanded in a plane-wave basis set with a kinetic energy cut-off of
8\,Ry. The Brillouin zone is sampled at special {\bf
k}-points,\cite{Monkhorst76} which are specified below for the different
systems when these are discussed.

In the following we describe in more detail the {\em damped Newton
dynamics\/} procedure to relax atoms, the Fermi surface smoothing
technique, and some technical improvements, which allow us to calculate
large systems.  We also describe the geometry of the systems
investigated and estimate the numerical accuracy of our results.

\subsection{Atomic relaxations}

In our adsorption calculations we typically allow the Al adsorbate and
the top two (111) layers or three (110) layers to relax until all force
components are smaller in magnitude than 0.04\,eV/\AA{}. It was checked
that relaxation of an additional layer leaves the adsorption-energy
differences practically unchanged.  The most important effect of the
adsorbate-induced substrate relaxation is a reduction of barriers for
bridge diffusion by 0.07\,$\pm$\,0.05\,eV.

The algorithm to relax the atomic geometry is based on {\em damped
Newton dynamics\/}. In a finite-difference form the time evolution of
any atomic coordinate $X$ is given by
  \begin{equation}
  X^{\tau +1} = X^\tau + \eta_X (X^\tau - X^{\tau - 1}) + \delta_X F^\tau_X\;,
\end{equation}
  where $X^{\tau}$ is the coordinate at time step $\tau$ and $F_X^\tau$
the force on $X$ at time step $\tau$.  The parameters $\eta_X$ and
$\delta_X$ control the damping and the mass of the coordinate.  The
choice of those parameters is guided by the goal that
this classical dynamics combines
a fast movement of the atoms
towards the next local minimum of the Born-Oppenheimer surface
and avoids  oscillations
around it.  We got fast convergence  for our Al surface systems with
$\eta_X \approx 0.6$ and $\delta_X\approx 8$.  This choice
brings the calculations
 close to the aperiodic limit of a damped oscillator
in classical mechanics.  Increasing the damping coefficient $\eta_X$
improves the stability of the  atomic relaxation process,
reducing it allows
for energy barriers to be overcome and so to escape from local minima.
In it's usage of the knowledge of the history of displacements the
damped dynamics technique is similar to the conjugate-gradient
technique.\cite{Gillan89}

Obviously the atomic geometry converges faster when larger displacements
per time step are executed.  The magnitude of useful displacements is
restricted, however, by the efficiency with which the electronic
wavefunctions converge to the electronic ground state of the new atomic
coordinates after the displacement. We found it advantageous to
have about eight purely electronic iterations after any atomic
displacement. Note that the time-consuming calculation of the atomic
forces is not done in those purely electronic iterations.
For all systems studied in this paper about ten atomic relaxations are
necessary to converge to the desired accuracy.

\subsection{Fermi occupation}

In order to stabilize the self-consistent calculations
for the electrons and to improve {\bf k}-space integration, we smear out the
Fermi surface. For this purpose the Kohn-Sham eigenstates of energy
$\epsilon_i$ are occupied according to a Fermi distribution $f=
f(\epsilon_i, T^{\rm el})$ with $k_BT^{\rm el} = 0.1$\,eV.  Using the
Fermi distribution implies that the free energy $F = E -T^{\rm el}S$ at
the electronic temperature $T^{\rm el}$ is minimized instead of the
total energy $E$.\cite{Gillan89,Neugebauer92,Vita91,Kresse93} Here $S$ is the
entropy of independent electrons,\cite{Kittel}
 \begin{equation}
  S = - 2 k_B \sum_i\left[f_i \ln f_i + (1-f_i) \ln (1-f_i) \right] \;.
\end{equation}
  This approach may cause some inaccuracies, as we want to get results
belonging to $T^{\rm el}=0$. For the free energy at a given geometry the
$T^{\rm el}\rightarrow 0$ limit can be easily obtained by evaluation of
$E^{\rm zero} = 0.5 (E+F) = E - 0.5 T^{\rm
el}S$.\cite{Gillan89,Neugebauer92,Vita91} This value differs from
$F(T^{\rm el}\rightarrow 0)$ only by terms which are third and
higher order in
$T^{\rm el}$.  For the optimization of the geometry the force $\partial
E^{\rm zero}/\partial X$ should be used which is, however, more
complicated to evaluate.\cite{MP} For our choice of $k_BT^{\rm el} = 0.1$\,eV
the geometries and the total-energy differences are almost not affected.
This was tested for the adsorption of Al on Al surfaces by using values
of 0.05\,eV and 0.2\,eV for $k_BT^{\rm el}$ and an increased number of
{\bf k}-points.


A further approach to stabilize the way
self-consistency is achieved is to
reduce electron transfer between  single-particle states in
successive iterations.
For this purpose fictitious eigenvalues after Pederson and
Jackson~\cite{Pederson91} are introduced.  The occupation numbers
are calculated directly from the fictitious eigenvalues according to
Fermi occupation at $T^{\rm el}$. These fictitious eigenvalues follow
the as-calculated eigenvalues in a sort of damped dynamics, so that both
sets of eigenvalues will become identical when self-consistency is
attained.  This indirect approach of damping charge transfer
oscillations is easier to implement than the more obvious
one of damping the change in occupation numbers directly.  The
reason is that the occupation numbers are constrained to be in
the range between 0 and 2, and their sum has to give the total number of
electrons. For the eigenvalues no such constraints exist.

\subsection{Optimizations}

The computer code is optimized for large systems by using the following
concepts.  Because the evaluation of the non-local part of the
pseudopotential dominates the computation time for large systems
 we utilize the translational symmetry for those atoms sitting on ideal
lattice sites.~\cite{Stumpf94b} Without introducing any approximation
this optimization typically reduces the number of operations by a factor
of ten for these atoms.

One often encountered problem with large systems is the $1/G^2$
dependence of the electrostatic potential. Here $\bf G$ is a
reciprocal-lattice vector.  This dependence leads to long-wavelength
charge-density oscillations, known as {\em charge sloshing\/} or {\em
$1/G^2$ instability\/}. See for example
Ref.\,\onlinecite{Arias92}.
We deal with this problem by
starting with a rather good initial density constructed by a
superposition of contracted atomic charge densities.  The contraction
was done following Finnis,\cite{Finnis90} where the radial atomic
densities are multiplied by a Fermi function.  The contraction
anticipates most of the intra-atomic charge transfer that occurs upon
building a solid from isolated atoms. The wavefunctions for the first
step of the self-consistent iterations are obtained by diagonalizing of
the Kohn-Sham Hamiltonian constructed from this approximate density and
within a reduced plane-wave basis ($E^{\rm cut} = 1.5 - 5$\,Ry,
depending on time and memory constraints). Then, in the first $\hat{\tau} < 8$
electronic iterations, the charge density $n({\bf r})$ is linearly mixed
like in ``standard'' self-consistent calculations:
  \begin{equation}
  n^{\rm in, \hat{\tau}+1}({\bf r}) = \alpha n^{\rm out, \hat{\tau}}({\bf r}) +
(1-\alpha)
  n^{\rm in, \hat{\tau}}({\bf r})\;.
\end{equation}
  We use a mixing coefficient $\alpha$ which increases from 10\,\% to
100\,\% within these first eight time steps. By this procedure the
charge sloshing was not initiated for the systems considered in this
paper and so did not cause problems during the rest of the calculations.
In calculations of larger cells than those reported here
we found the linear mixing in {\bf r}-space insufficient. The charge density
sloshing could however be efficiently suppressed by a mixing in
{\bf G}-space
with a mixing coefficient $\alpha(G)$
which was chosen smaller for smaller $G$.

For large systems the required computer memory raises as the square of
the number of atoms and it is largely determined by the number of
wavefunction coefficients. We optimize memory usage in several ways. A
simple steepest-descent update procedure for the wavefunctions is
used\cite{conjgrad} so that only the wavefunction coefficients of one
iteration need to be stored.  The wavefunction coefficients and most of
the other large arrays are stored in single precision; however, double
precision is used for all floating point operations and for storing
intermediate results.

Our  computer code  optimizes the data access in computers
which use memory of different speed.
The idea is that once data is transferred from slow memory to fast
memory, e.\,g.~from the disk to main memory, this data should be used as
often as possible before it is moved back to the disk.  This improvement
can be accomplished by reordering loops or by blocking
techniques.\cite{Stumpf94b} The most important case where blocking is
used is the orthogonalization of the wavefunctions.  Instead of
orthogonalizing always one wavefunction to those with lower index (the
standard Gram-Schmidt procedure), we orthogonalize a block of, say, 30
wavefunctions to those with lower index and then orthogonalize the
wavefunctions within the block.  In the case where only part of the
wavefunctions at one {\bf k}-point fit into main memory this procedure
reduces the disk to memory data transfer by a factor up to 30.

An example of the efficiency of the code is the calculation of an Al
slab with 350 atom per unit cell in a supercell as large as 560 atomic
volumes, and sampling the Brillouin-zone at one special {\bf k}-point.
This leads to 28\,000 plane waves, to 560 involved
electronic states, and an overall memory
requirement of 200\,MB. This calculation takes about 25\,h
on an IBM/6000~370 RISC workstation with 64\,MB main memory, if all atoms sit
on ideal lattice positions. If no atoms are sitting on ideal sites, the
time increases by a factor of three.  The time spent waiting for disk
access during the calculation is below 30\,\% and could even be reduced
noticeably by using the faster hard disks available today.

\subsection{Slab geometry}

In order to describe an adatom on a crystal surface we use a slab in a
supercell.  The repeating slabs are isolated by $\gtrsim 8$\AA{} of
vacuum spacing.  For the purpose of studying practically isolated
adsorbates, the distance between adatoms in neighboring cells is at
least three nearest-neighbor spacings.  This results in an interaction
energy below 0.03\,eV for Al adatoms.

In order to have more bulk like layers and to avoid artificial
adsorbate-adsorbate interaction through the slab, we adsorb the Al
adatom only on one side.  This approach reduces the slab thickness
necessary for the desired degree of accuracy.~\cite{Neugebauer92} Due to
the unsymmetrical situation an artificial electric field perpendicular
to the slab might arise.  This field is compensated in our calculations
as described in Ref.\,\onlinecite{Neugebauer92}, by introducing a dipole
layer in the vacuum region. In the case of an Al adsorbate on an Al
surface this field is always very small so that even in the uncompensated
case the energy differences between different sites are practically
unaffected.

For the calculation of adsorption on Al\,(111) we use
typically normally 5-layers-slabs.
Calculations with slabs of \mbox{4,} \mbox{6,} and 7
layers show that even with a 4~layer slab adsorption energy {\em
differences} are accurately given, which means that they change by less
than 0.03\,eV when thicker slabs are used.

For Al\,(100) we find that the desired accuracy of 0.03\,eV requires a
slab thickness of at least 6~layers. The quantity most
sensitive to the slab thickness is the energy barrier for
exchange diffusion; for a 5 layer slab this is by
0.25\,eV or 66\,\% lower
than that of the
6-~and 7-layers thick slabs (see Appendix \ref{Al100}). We use a
4\,$\times$\,4 surface cell for the calculations of
self-diffusion on Al\,(100). For the
Al\,(110) surface we used 8~layers and a 3\,$\times$\,4 surface cell.

In this paper we treat the two densely packed steps on Al\,(111). One is
called \{111\}-faceted, the other \{100\}-faceted (see
Fig.~\ref{StepScheme}).  The \{111\} and \{100\} micro-facets are the
steepest ones and therefore give an unambiguous way of naming the
steps.\cite{Lang72} We shall see, however, that the electronic
properties of the \{111\}-faceted step are more closely related to the
(110)-surface. This similarity was already discovered by Nelson and
Feibelman,~\cite{Nelson92} who showed that the relaxation of the
Al\,(110) surface atoms and the atoms close to the $\langle
110\rangle/$\{111\} step is very similar and that the charge densities
at the Al\,(110) surface and along the \{110\} plane trough a $\langle
110\rangle/$\{111\} step are nearly identical. Our calculations confirm
this and also show the similarity of chemical properties and of
self-diffusion on the Al\,(110) surface and at the \{111\}-faceted step.

In order to analyze the step properties we use three different
approaches.  In the first approach a slab in (111) orientation is put in
an orthorhombic supercell.  Then half of the atoms of the top layer are
removed, so that the remaining grooved surface has two steps, one being
\{111\}- and the other \{100\}-faceted (see Fig.~\ref{StepScheme}; the
lower part of the supercell, containing the flat termination
of the slab and some more vacuum, is omitted in the upper
panel of Fig.~\ref{StepScheme}).  We choose different sizes of
the surface supercell to study the influence of finite size
effects. The width of the cell in $[1\bar{1}0]$ direction is
varied from three to four atoms and the width of the terrace
in $[11\bar{2}]$ direction is three to four atomic rows. All
these systems give results which differ only by $\leq
0.05$\,eV.  One reason for this is the rapid screening of Al.
The other reason is that quantum size-effects are often
unimportant for total-energy differences on stepped Al(111)
which was already noted by Nelson and
Feibelman.~\cite{Nelson92,Feibelman92}

The second approach uses a class of systems which are slabs of
$(m,m,m-2)$ and of $(m+2,m,m)$ orientation where $m$ is an
even number.  Surfaces with these Miller indices are vicinal
to the (111) surface.  The $(m,m,m-2)$ surface consist of
terraces of (111) orientation which are $m$ atomic rows wide
and separated by $\{11\bar{1}\}$-faceted steps.  The
$(m+2,m,m)$ surface has (111) terraces $m+1$ atomic rows wide,
which are separated by \{100\}-faceted steps.\cite{Lang72} The
relationship between the Miller indices of the vicinal
surfaces and the constituent low index facets can be seen by
the vector decompositions $(m,m,m-2) = (m-1)\times(111) +
1\times(11\bar{1})$ and $(m+2,m,m) = m\times(111) +
1\times(200)$. Note that conventionally common factors are
removed from Miller indices, so that instead of (200) [which
is the shortest reciprocal lattice vector in the (100)
direction] the more familiar (100) is used and for even $m$
the common factor two is removed.  Thus the Miller indices are
$1/2\times(m,m,m-2)$ and $1/2\times(m+2,m,m)$.

We worked with surfaces belonging to even $m$ because they can be
accommodated in a monoclinic supercell, whereas for odd $m$ a triclinic
supercell is required.  The first surface (i.\,e.~$m=2$) of the
$(m,m,m-2)$ family is the (220)$\equiv$(110) surface, for which the (111)
terraces are so narrow that no surface atom has a (111)-like
coordination.  After some test calculations with the (221)
surface ($m=4$), we
concentrated on the (332) surface ($m=6$)
for studying the properties of nearly
isolated \{111\}-faceted steps. The (332) surface has (111) terraces
which are six atomic rows wide (see Fig.~\ref{Systeme1}).  We used a
1\,$\times$\,4~surface unit cell, which means that 6\,$\times$\,4~atoms
are exposed at the surface.  This layer is repeated six times to build a
slab containing 144~atoms per cell.

For studying the properties of nearly isolated \{100\}-faceted
steps we used the (433) surface out of the $(m+2,m,m)$ family,
which contains seven atomic rows of (111)-orientation. With a
1\,$\times$\,4 surface unit cell, we get 7\,$\times$\,4 atoms
exposed on each surface (see Fig.~\ref{Systeme2}). Again,
6~layers were taken which gives a slab containing 168~atoms
per cell.

The adsorption calculations at the steps of the (221), (332), and (433)
surfaces essentially reproduce the results of the grooved surfaces,
which reflects the efficient screening at Al surfaces. This
agreement of results obtained for different slabs also provides a good
test of the convergence of our calculations with respect to system size
and {\bf k}-space sampling.

The main advantage of using the vicinal surface systems is that they
allow the investigation of long-range adsorbate-step interactions, which
were found for example experimentally by Wang and Ehrlich;\cite{Wang93}
for a given adsorbate-step distance the number of atoms in the cell is
only slightly more than half of that required for the grooved slab
geometry.

For the purpose of calculating the difference in step formation energy
of \{111\}- and \{100\}-faceted steps a third kind of system has to be
used.  It should contain only one kind of step and it should provide a
favorable error cancellation when the energy {\em difference} of the
\{111\}- and \{100\}-step systems is determined. The chosen system consists
of a triangular Al island on Al\,(111) with either \{111\}- or
\{100\}-faceted edges, depending on the triangle's orientation.  Note
that this way the number of atoms and the {\bf k}-space integration is
the same for both step orientations.  One new uncertainty in determining
the step formation energy difference is now introduced however: also the
corners of the triangles contribute to the difference. The contribution
of the corners has to be removed by extrapolating to the thermodynamic
limit of infinite island size.  The largest triangles studied consisted
of 21 atoms in a $8\times7$ Al\,(111) surface cell.  At a slab thickness
of 4 layers this gives 245 atoms per supercell.

\subsection{Estimated accuracy}

The equilibrium lattice constant for a basis set with 8\,Ry cut-off and a
{\bf k}-point sampling like that used for the surface calculations is
3.98\,\AA{} (including zero-point vibrations),\cite{Neugebauer92} which
is 0.7\,\% smaller than the experimental value.  The cohesive energy is
4.15\,eV, which is 0.75\,eV higher than the experimental one of
3.40\,eV.~\cite{cohesive} These errors in bulk results are within the
expectations for a well-converged DFT-LDA calculation.

The 8\,Ry plane wave cut-off was tested to be sufficient to converge
adsorption energy {\em differences\/} to a numerical accuracy better
than 0.02\,eV (see also Ref.\,\onlinecite{Neugebauer92}).  Because of
the large size of our supercells -- they comprise, depending on the
problem, 140\,--\,560 atomic volumes -- one special {\bf
k}-point~\cite{Monkhorst76} in the irreducible quarter of the surface
Brillouin zone, which is always nearly square, is sufficient to give
energy differences which are within 0.03\,eV of those obtained by using
two or four times the number of {\bf k}-points.

We also tested the dependence of our results on system size.  Here the
slab thickness as well as the adsorbate-adsorbate and the step-step
interactions are relevant.  System size and {\bf k}-space-sampling
effects are difficult to separate, because often a change of the size of
the system implies different {\bf k}-sampling, and the two effects are
about equal in magnitude. We therefore cannot quantify the error
introduced by system size effects separately.  In order to reduce errors
from these two sources we always quote the mean value of calculations at
different {\bf k}-point sampling and system size. This improves the
accuracy because the variations with system size and {\bf k}-space
sampling are often oscillatory. We obtain an overall numerical accuracy
of the energy differences given of $\leq 0.06$\,eV, unless a different
error margin is stated explicitly.

\section{Differences of ideal \{111\}~ and \{100\}~faceted steps}
\label{Differ}

Calculations of the average energy of the two step types can
be performed most accurately by investigating the total energy
of surfaces with terrace stripes (see Fig.~\ref{StepScheme}),
and comparing these results to those of the flat surface. The
mean step formation energy is determined as 0.24\,eV per step
atom. Table~\ref{SurVacForm} shows that this is about half of
the energy required to create the Al\,(111) surface per
surface atom, and that the step formation energy compares to
the difference of the surface energies between Al\,(111) and
the rougher (100) or (110) surfaces per atom.



The energy difference of the two step types can be obtained by investigating
triangular islands adsorbed on Al\,(111), as these contain only one type
of steps (see Fig.\,\ref{Triangle}).  Comparing islands with 6, 10, 15,
and 21 atoms we can extrapolate to the limit where the influence of the
corner atoms is negligible.  Table~\ref{TriaDat} lists the results for
the total-energy differences of two triangles whose orientations differ
by~$60^\circ$, and hence have different step types. The data show the
rapid convergence of this energy difference with island size.  Dividing
these energy differences into an island size independent contribution
from the three corner atoms and a contribution proportional to the
number of true edge atoms a fit to our data (see Table~\ref{TriaDat})
gives the following results.  Triangular islands with \{111\}-faceted
steps are more favorable by 0.025~eV per corner and by 0.017~eV per true
step atom than islands bounded by \{100\}-faceted steps.  The energy
differences are almost the same, whether the island atoms are relaxed or
not.  This small effect of relaxation shows that the step formation
energy difference is an electronic effect and is not determined by a
different step-induced atomic relaxation.
   It is interesting that our results cannot be estimated from simple
embedded atom or effective medium theory\cite{Liu92,Nelson93,embedding} or a
bond-cutting model.\cite{Bondcut} The reason is that the two different
triangular islands have exactly the same number of bonds.

 From the step formation energy one can directly obtain the equilibrium
shape of larger islands by minimizing the island free energy for a given
number of atoms in the island. The graphical solution of this
optimization problem is called the Wulff construction.\cite{Zangwill88}
As a result one obtains hexagonally-shaped islands, where the edges
alternate between those with a \{100\} and those with a \{111\}
micro-facet. The calculations predict that in thermodynamic equilibrium
the \{111\}-faceted edge should be longer with a edge length ratio
$L^{\langle 110\rangle/\{100\}}:L^{\langle 110\rangle/\{111\}}$ of $4:5$.
Effects of the vibrational or the configurational entropy on the length
ratio, which might be important at higher temperatures, are neglected.

It is interesting to note that such hexagonal islands have been observed
experimentally by Michely and Comsa~\cite{Michely91} in their STM
studies of growth and sputter removal of Pt\,(111).  These experiments
show that the \{111\} micro-facet is favored which is what we predict
for Al\,(111).  There is a quantitative difference, since for Pt\,(111)
the measured edge-length ratio is 0.66, i.\,e.~$2:3$. The
similarity to our results is much more than what one would have
expected, as, in general, Al and Pt behave quite differently.

\section{Work function differences and induced surface dipole moments}
\label{IndDip}

Our calculations of induced surface dipole moments and
work function differences at Al surfaces give unexpected results.

The first interesting observation is that the Al\,(111)
surface and the Al\,(110) surface have about the
same work function $\Phi$.  Our best converged results
are $\Phi_{\rm Al\,(111)} = \Phi_{\rm
Al\,(110)} = 4.25$\,eV,\cite{averlay} the experiment finds $\Phi_{\rm
Al\,(111)}^{\rm exp} = 4.24$\,eV and $\Phi_{\rm
Al\,(110)}^{\rm exp} = 4.28\,$eV.\cite{Grepstad76}

The second result is that steps on Al\,(111) do not affect the
work function very much.  Table~\ref{TabDip} lists the induced
dipole moment $\mu$ per step atom.  The \{111\}-faceted step
induces practically no dipole ($\mu$=-0.01\,debye), the
\{100\}-faceted step has a small dipole moment with the
negative end pointing into the vacuum ($\mu$=0.045\,debye),
which means that they increase the work function.  Induced
dipole moments translate into work function changes $\Delta\Phi$ according
to the Helmholtz equation
\begin{equation}
  \Delta\Phi = 135.\frac{\mu}{A}
\end{equation}
with $\mu$~in debye, $\Delta\Phi$~in eV, and the area $A$ per
dipole in bohr$^2$.

More noticeable dipole moments are found for 3-fold
coordinated Al adatoms (see Table~\ref{TabDip}) An example is
the dipole moment of 0.24\,debye for an Al adatom on the
hcp-site. If there was an adlayer of those Al adatoms on
Al\,(111) of, say, 1/10 monolayer coverage, the work function would
increase by 0.13\,eV.



The reported results on induced dipole moments and work
function differences contradict the traditional model of
charge redistribution at rough metal surfaces and around
protrusions on metal surfaces like steps or
adatoms.\cite{Hoelzl79,Zangwill88} This model is based on
Smoluchowski smoothing.  Smoluchowski smoothing is driven by
the kinetic energy of the electrons which is lower for a less
corrugated charge density.  The smoothing of the charge
density lowers the work function for rougher surfaces and
causes a induced dipole moment with the positive end towards
the vacuum for protrusions on the surface.  The smoothing
effect is often discussed in a nearly free electron picture.
An example is the calculation by Ishida and
Liebsch\cite{Ishida92} of the induced dipole moment of steps
on jellium. They find that steps reduce the work function.
Extrapolating their results for a step on Al\,(111) one gets
an induced dipole moment of about~$-0.07$\,debye per step atom
equivalent.

The reason why the smoothing model fails in our calculation
of atomic Al could be that the smoothing effect, which is
certainly there, is (over-)
compensated by the attraction of electrons towards the deeper
potential around surface atoms on Al(110), step-edge atoms on
Al\,(111), or adatoms on Al\,(111). These atoms are only 7- or
3-fold coordinated as compared to the 9-fold coordinated
surface atoms.  Therefore they are less well screened, which
leads to the deeper potential in their vicinity.  This effects
a net transfer of electrons towards those undercoordinated
atoms. A more thorough discussion of the charge rearrangements in
different surface geometries and the resulting induced dipole
moments will be published elsewhere.\cite{Ari94}


Having a possible explanation why the standard model fails in
the case of the simple metal Al the remaining puzzle is why it
seems to work for the transition metals.\cite{Hoelzl79}
To give an example. Steps
on Au\,(111) and Pt\,(111) show dipole moments between --0.25~debye (Au)
and --0.6~debye (Pt) per step atom.\cite{Besocke77} A comparison with
the jellium calculations in Ref.\,\onlinecite{Ishida92} shows that
for those steps the induced dipole moment is larger in
magnitude than it would be expected from the smoothing effect
of the $s$-$p$ like electrons only. The
additional negative dipole moment is likely caused by a
polarization of the $d$-electrons of the step atoms.  This would also
explain why Au shows a smaller effect than Pt. Au has a filled
$d$-shell, in Pt the Fermi level cuts the $d$-band and
therefore it is easier to polarize the $d$-states.

We see that for the simple as well as for the transition
metals important additions to the smoothing based model for
induced surface dipole moments and work function differences
have to be made.

\section{Al adatoms on flat and stepped Al\,(111)}
\label{Al111}

This section describes the total-energy surface for an Al
adsorbate atom on the flat Al\,(111) surface and at the
\{100\}- and \{111\}-faceted steps.  This discussion is
directly relevant for surface diffusion and crystal growth on
Al\,(111). We will study how an Al adatom moves across the
Al\,(111) surface, what happens when the adatom comes close to
a step, how it attaches to the step coming from the lower
side, and how it incorporates into the step by an atomic
replacement process coming from the upper side.


\subsection{Diffusion on flat Al\,(111)}

The diffusion energy barriers of an isolated Al adatoms on the flat
Al\,(111) surface is very small (0.04\,eV; see Table~\ref{D111} and
Figs.\,\ref{Diff111S} and~\ref{Diff100S}). The hcp site is the stable
binding site and the energies of bridge and fcc sites are almost
degenerate.  The diffusion path between the hcp sites is the direct
connection between adjacent hcp, bridge, and fcc sites.

The only marked maximum of the total-energy surface is at the atop site.
The top site is 0.53\,eV higher in energy than the hcp site. Interesting
about the atop site is the low height of the adatom at the atop site
which is close to the height in the 3-fold sites (see Table~\ref{D111}).
This is a consequence of the fact that bond length gets smaller when the
coordination is lower (for the atop site we obtain a bond length of
2.51\,\AA{} which is 6\,\% smaller than for the 3-fold sites).
Furthermore, we find that the adatom at the atop site introduces a
strong substrate relaxation: the substrate atom below the adsorbate is
lowered by 0.4\,\AA{}. A similar substrate relaxation was found in
calculations for alkali-metal adsorbates on
Al\,(111).\cite{Neugebauer92,Stampfl94}

The height of the diffusion energy barrier of 0.04\,eV is at the lower
limit of an energy barrier one would expect for this close-packed
surface from experimental data (see Table~\ref{Tab1}). Two features of
the total-energy surface are worth discussing further.

We first note that it is unexpected that the hcp site is the equilibrium
site for low coverage.  The hcp and fcc sites both provide threefold
coordination, but only the fcc site continues the ABCABC stacking of the
fcc crystal, whereas the hcp site belongs to an ABCAC stacking. A simple
continuation of the fcc structure would suggest that the fcc site should
be occupied.  In fact, the fcc-hcp energy difference is coverage
dependent. Above 1/4~ML coverage the fcc site gets more stable.  To
create a full monolayer of Al at the hcp position costs 0.05\,eV per
surface atom compared to the fcc stacking.  It is interesting that this
energy is exactly the mean of the formation energy of the three bulk
stacking faults in the $\langle111\rangle$-direction, which were
calculated by Hammer et al.\cite{Hammer92}

The reason for the different adsorption energy at the hcp and fcc sites
at low coverage is not obvious. Half of the difference exists already
before the Al\,(111) substrate is relaxed, which shows that there is an
electronic contribution. Another hint for an understanding might be that
the Al adsorbate at the fcc site has a larger induced dipole moment
than at the hcp site (see Table~\ref{TabDip}).

The second interesting feature of Al on Al\,(111) is that the three-fold
coordinated fcc sites and the two-fold coordinated bridge sites have
practically the same total energy. This is not explainable in a
coordination number model.\cite{Bondcut} One might guess that the rather
favorable energy of the bridge site is a result of the substrate
relaxation. Indeed, on the unrelaxed substrate the bridge site is
energetically less favorable (but only by 0.07\,eV) than the fcc site.
If the atomic geometry were not relaxed and all atoms are kept at bulk
nearest neighbor spacings, then the fcc site were favored by 0.13\,eV
over the bridge site. Still, this is much less than what a coordination
number model would predict.

The energy difference between hcp, bridge, and fcc sites is very small.
It is therefore important to check, if the calculations are sufficiently
accurate.  We therefore performed several test calculations, varying
carefully all parameters which affect the accuracy. We used coverages
from 1/12~ to~1/56 ML, increased the number of {\bf k}-points from~1
to~4 and to~9, we used from~4 to~7 Al\,(111) layers, and we increased
the plane wave cut-off. The energy differences of different sites were
very stable and the order of fcc and hcp site never reversed at low
coverage.  We expect the difference between hcp and fcc site to be
accurate to within 0.02\,eV. The relative accuracy for the bridge site is
slightly worse. Feibelman's recent LDA
calculations\cite{Feibelman92}
found the hcp site more
stable than the fcc site by 0.03\,eV, in good agreement with our
results.

In order to get an idea of the nucleation probability and the influence
of adsorbate-adsorbate interaction on the mobility we calculated the
adsorbate-adsorbate interaction between two Al adatoms sitting at
neighboring hcp-sites. The energy gain with respect to isolated adatoms
is 0.58\,eV.
We note, however, that only one configuration (hcp-hcp) was considered
and we did not perform any of our usual checks, so that here an error of
$\pm$\,0.2\,eV might be possible.

\subsection{Approaching the step}

Table~\ref{D111} and Figs.~\ref{Diff111S} and~\ref{Diff100S} show that
the Al adatom is attracted by the step on the lower as well as on the
upper terrace and that this is similar for both sorts of steps. This
attraction leads to an energy gain of about 0.1\,eV at the threefold
sites directly at the upper step edge compared to the flat Al\,(111)
surface. The long-range attraction towards the step as experienced by
atoms on the lower terrace is weaker, but at closer distance the
attraction gets as strong that the last two threefold sites before the
fivefold at-step site are not local minima any more and an adsorbate
there will be funneled towards the step directly.

This attraction and the funneling of Al adsorbates by the lower step
edge resembles the behavior found recently by Wang and Ehrlich for Ir on
Ir\,(111).\cite{Wang93} Also on Pt\,(111) indications of a funneling
towards steps exist.\cite{Bott92} Finding the same behavior for
different metals suggests that the medium-range to long-range attractive
interaction is a more general phenomenon. Therefore, an understanding of
the nature of the mechanism behind this attraction of adatoms towards
steps is desirable. We approach the analysis of its origin by
considering the following thinkable  mechanisms and check
whether they are indeed attractive or not.  The mechanisms are
dipole-dipole interaction, elastic interaction,
and interaction of defect induced surface-states.

The analysis shows that the mechanism is not an attractive dipole-dipole
interaction.  The dipole-dipole interaction in the studied cases will be
very weak. Even for the largest dipole moments (those for adatoms at the
fcc site and for the \{100\} step, see Table~\ref{TabDip}) the
interaction energy would be below 1\,meV for distances larger than one
nearest-neighbor spacing.  Furthermore we note that
the interaction would be
{\em repulsive\/}.

Another possible mechanism responsible for the attractive
adatom-step interaction could be the elastic interaction of
the adsorbate-induced and the step-induced relaxation fields.
This possibility can easily be checked by a calculation of the
energy while keeping the substrate atoms at their ideal
positions. Our calculations show that the long-range
interaction remains unchanged. Thus, elastic effects can be
disregarded as the origin of the attractive interaction
between the adatom and the step. The fact that the adatom-step
attraction is not of elastic origin and that it is long range
excludes the possibility that it can be reproduced with more
simple bonding models like coordination number models\cite{Bondcut} or
effective medium and embedded atom.\cite{Daw84,Jacobsen87}

In consequence there remains only the possibility that the attractive
interaction is caused by an interaction of adatom-induced and
step-induced surface states or screening charge densities. This
conclusion is supported by recent STM pictures of adsorbate and step
induced surface states on Cu\,(111) and
Au\,(111).\cite{Crommie93,Hasegawa93}

\subsection{Comparison of self-diffusion on the flat Al\,(110) surface and at
  the \{111\}-faceted step on Al\,(111)}

Figure~\ref{Kugeln} shows five geometries which we consider to be
particularly important for diffusion on the Al\,(110) surface and at the
\{111\}-faceted step on the Al\,(111) surface. In the following we will
compare the two systems and we will analyze the factors determining the
adsorption energies in different bonding geometries.  The results of
Table~\ref{D110} show that at geometries having the same
nearest-neighbor environment, the adsorption energies are very similar
on the (110) surface and at the \{111\}-faceted step.

At the two fivefold sites (Fig.~\ref{Kugeln}\,(a)\,) the adsorption
energy is rather large and practically identical in both cases.  The
calculated Al bulk cohesive energy is only 7\,\% or 0.27\,eV larger. The
adsorption energy on the threefold equilibrium site on Al\,(111) at low
coverage (the hcp site) is as much as 20\,\% or 0.78\,eV smaller
(Table~\ref{D111}). A coordination number
model\cite{Bondcut} would give too small an estimate of the
adsorption energy. The high at-step adsorption energy we
understand as
follows. At both fivefold sites the adsorbates sit in a valley
which should be filled
with additional electronic charge due to
Smoluchowski-smoothing.\cite{Smoluchowski41} Additionally
the adsorbates sit very close to the surface.  On the (110) surface the
height at the fivefold site is 1.33\,\AA{} relative to the relaxed clean
surface. This should be compared to the hcp site on Al\,(111), where the
height of the adsorbate is 2.09 \AA. The combination of increased charge
density and low height provides a good embedding of the
adsorbate.\cite{embedding}

%
%


\subsection{Adatom at bridge sites}

The twofold coordinated bridge sites are possible saddle point configurations
for surface self-diffusion.
  The energies of the comparable bridge sites on Al\,(110) and at the
\{111\}-faceted step are similar. However, the two different bridges
(see Figs.\,\ref{Kugeln}\,(b) and \ref{Kugeln}\,(c)\,) have a markedly
different energy. This energy difference gives rise to a difference in
barrier height of a factor of two (see Table~\ref{D110}).  What is the
reason for this difference of sites with the same coordination\,? Our
explanation is the same as for the high adsorption energy at the 5-fold
site.  The higher adsorption energy for the long bridge is accompanied
by a lower distance to the surface.  For the long bridge on the (110)
surface the adatom height is only 1.58~\AA{}, whereas the height is
2.16~\AA{} for the short bridge.  Another effect should be
that
Smoluchowski-smoothing  fills the valleys with electron density taken
from the upper part of the rows on the clean surface.  For the
short-bridge position, this would reduce the charge density around the
adsorbate, while it would rather increase the embedding charge density for the
long bridge position.

Having seen the striking similarity of adsorption energies at
the (110) surface and at the \{111\}-faceted step, it is no
surprise that a system which lies between the two cases,
namely the (331) surface, shows a very similar
behavior.\cite{Feibelman92} For this system Feibelman obtained
energy differences between the Al adsorption at 2-fold and
5-fold sites which are very close to ours (see
Table~\ref{D110}). We expect that the agreement with our
results would be within 0.05\,eV if Feibelman would have
included the adsorbate-induced relaxation of the substrate.
The agreement of both studies is a most demanding test for the
numerical accuracy of both calculations, since Feibelman used
a rather different technique.

We will now show that the just discussed bridge sites are not
the lowest energy transition states for diffusion for the
surfaces considered.  Diffusion takes place preferably via
exchange mechanisms.

\subsection{Exchange process for diffusion parallel to
  $\langle110\rangle/\{111\}$ steps on Al\,(111) and along the channels of
Al\,(110)}


Investigating the Al\,(110) surface we  will first show that the exchange
mechanism lowers the energy barriers for diffusion, and this holds even
in the so called easy direction, i.\,e.\ for diffusion in the channels
of this surface.  Figure~\ref{Kugeln}\,(d) sketches the symmetric
configurations of the exchange paths. For both the (110) surface and for
the \{111\}-faceted step, the geometry may be described as two fivefold
coordinated Al adatoms which are close to a surface vacancy. At the
(110) surface the two atoms are displaced from the ideal fcc lattice
positions by 0.08\,\AA{} towards the vacancy and they are 0.05\,\AA{}
closer to the surface than for the ``normal'' fivefold site. At the step
the corresponding displacements are similar. To get some insight into
the nature of bonding of this configuration, we estimate its energy from
the adsorption energy of the ``normal'' fivefold site $E_{\rm{ad}}$ and
the calculated vacancy-formation energies, where the Al chemical
potential, which defines the energy of the removed Al atom, is taken as
the bulk cohesive energy, i.\,e.~$4.15$\,eV. For the vacancy formation
energy at the (110) surface we then obtain $0.12$\,eV and at the step
the result is $E^{\rm{vac}}_{\rm{step}} = 0.21$\,eV (see
Table~\ref{SurVacForm}).  The energy barriers are then estimated as
$E^{\rm{vac}}_{\rm{(110)}} - E^{\rm{ad}}_{\rm{(110)}} = 0.38$\,eV and
$E^{\rm{vac}}_{\rm{step}} - E^{\rm{ad}}_{\rm{step}} = 0.49$\,eV, in
close agreement with the full calculation of the exchange-parallel
processes in Table~\ref{D110}.

Up to now we have only discussed the symmetric configurations
of the exchange diffusion path (see Fig.\,\ref{Kugeln}\,(d)\,)
without knowing if these are really saddle point
configurations. Additional calculations at the step show that
they are {\em not\/} saddle points, but that they correspond
to very shallow minima of the total-energy surface. The saddle
points found in these calculations have an energy
about~0.04\,eV higher than the energy of the symmetric
configuration.  This energy difference is about 1/10 of the
total energy barrier and also very small compared to the
numerical accuracy of our calculations. It is therefore
appropriate to discuss the energy for the symmetric exchange
configuration when talking about diffusion barriers.

In addition to the discussed mechanisms a third diffusion mechanism for
diffusion along the channels was recently proposed by Liu et al.\cite{Liu91}
They argue that the configurations of Fig.\,\ref{Kugeln}\,(e), which are
saddle point configurations for the diffusion perpendicular to the channels,
are
also approximately saddle point configurations for the diffusion parallel to
them.  This  would imply that the diffusion parallel and the
exchange diffusion perpendicular to the rows would have about the same energy
barriers. This idea is interesting, but we get a higher energy barrier for the
perpendicular than for the parallel exchange process (see next
paragraph and Table~\ref{D110}).  Thus, for Al we cannot support the idea
\cite{Liu91} but nevertheless, the proposed process could be the
explanation for the near identity of parallel and
perpendicular to step diffusion
energy barriers on Ni, Ir, and Pt (see Table\,\ref{Tab1}).
\label{diff110cross}

\subsection{Exchange process for diffusion perpendicular
  to $\langle110\rangle/\{111\}$ steps on Al\,(111) and to the channels
of Al\,(110)}

We now discuss the exchange perpendicular to the channels of Al\,(110)
and across the $\langle110\rangle/\{111\}$ step on Al\,(111) (see
Fig.~\ref{Kugeln}\,(e) and~\ref{Diff111S}). Again the energies of the
saddle point configuration on Al\,(110) and at the step are similar (see
Table~\ref{D110}). The slightly higher energy of the saddle point
geometry at the step can be understood in terms of the higher vacancy
formation energy at the step (see Table~\ref{SurVacForm}) and of the
threefold rather than fourfold coordination of the upper Al atom. On
Al\,(110) the energy of the exchange configuration can be estimated
analogously to our discussion in the last Section. The
configuration consists of two neighboring fourfold coordinated atoms
close to a vacancy. We want to approximate the energy of the fourfold
coordinated atoms by that of another fourfold atom adsorbed at the same
height.  There is however some ambiguity in assigning a height to the
two adsorbates.  The possibilities are to take the distance of
0.80\,\AA{} to the (110) surface or that of 2.05\,\AA{} to the
\{111\}~micro-facets, which are built by forming a surface vacancy on
Al\,(110) (compare Fig.\,\ref{Kugeln}\,(e)\,). The height above the
\{111\}~micro-facets is more relevant here than that above the (110)
surface because all substrate nearest neighbors of the two
adsorbates belong to these \{111\}~micro-facets, and we showed
already that the nearest neighbor environment determines the
bonding to a large extent as a consequence of the good
screening of Al.  The adsorption energy can therefore be
approximated by $E^{\rm{ad}}_{\rm{(100)}} = 3.77$\,eV of an Al
adatom on the fourfold site on Al\,(100), which sits
1.69\,\AA{} above the surface (see Table~\ref{Diff100Dat} in
the appendix).  This gives an estimated energy barrier of
$E^{\rm{vac}}_{\rm{(110)}} + E^{\rm{ad}}_{\rm{(110)}} - 2
E^{\rm{ad}}_{\rm{(100)}} = 0.62$\,eV, which equals exactly the
energy barrier found in the full calculation.

The success of the simple approach used above to assemble the energy for the
exchange configurations
from the energies of its constituents exemplifies the role of height and local
environment for binding on Al surfaces. This also indicates that the nature of
the bonding is very similar for equilibrium adsorption sites and in the
examined exchange
configurations.

\subsection{Comparison of self-diffusion close to the
$\langle110\rangle/\{111\}$ and to the $\langle110\rangle/\{100\}$ step
on Al\,(111)}

The above studies showed that the interaction of an Al adatom with the
two kinds of close-packed steps on Al\,(111) is at larger distances
attractive and practically identical. Directly at the step however we
identified some important differences in adsorption energies and diffusion
barriers and mechanisms.

The results given in Table~\ref{D110} show that the adsorption energies
at the fivefold coordinated at-step sites is nearly the same, but that there
is a small preference (0.03\,eV) for the $\langle110\rangle/\{100\}$
step. This energy difference is very small; however it might obey a
general rule and is therefore interesting to analyze.  According to
Nelson et al.,\cite{Nelson93} it can be explained by the different step
formation energies (see Table~\ref{SurVacForm}).  Adsorbing, e.\,g., an
Al atom at the \{100\}-faceted step creates two
\{111\}-faceted ``micro-steps'' (The situation is reversed at the
\{111\} step). The creation of \{111\}-micro-steps should be
favorable, because \{111\}-faceted steps are favorable. According to
this model, the adsorption energies should differ by $2\times
0.017$\,eV. Indeed, the calculated at step adsorption energy difference
is very close to this value.

It is tempting to apply the same type of discussion also to the creation
of vacancies within the step. To form a vacancy at the \{100\}-faceted
step creates \{111\}-micro-steps and should be favorable. However, the
full calculations show that such analysis is not appropriate; the
vacancy formation energy at the \{111\}-faceted step is energetically
favorable (see in Table~\ref{SurVacForm}).

\subsubsection{Diffusion along the step}

The results of Table~\ref{D110} imply that an Al adatom diffusing along
the \{100\}~step will use the normal bridge hopping mechanism and not
the exchange mechanism as at the (100) surface or at the
\{111\}~step. The main reason for this difference between the two steps
is that the adsorption energy at the long-bridge position at the
\{100\}~step is 0.19\,eV higher. We understand this
as an effect of geometry (see, e.\,g., Figs.~\ref{Diff100S} and
\ref{Kugeln}\,(b)\,).
Whereas at the long-bridge position of the \{111\} step the adsorbate is
clearly twofold coordinated, the coordination is nearly fourfold at
long-bridge position at the \{100\} step. The exchange-parallel
configurations at the \{111\} step and at the \{100\} step have
practically the same energies. As a consequence it follows that the
diffusion at the lower terrace parallel to the \{100\} step proceeds via
hopping mechanism with an energy barrier about 0.1\,eV lower than that
for the exchange diffusion along the \{111\} step. The different
diffusion mechanisms should lead to different diffusion
prefactors~$D_0$. The impact of this difference for the temperature
dependence of crystal growth is discussed below in Section VI.

\subsubsection{Diffusion across the step}

For diffusion across the \{111\} and the \{100\}~step we obtain in both
cases exchange diffusion mechanisms with nearly the same energy barrier
heights (see Figs.~\ref{Diff111S} and~\ref{Diff100S}).  For the
diffusion across the step in the descending direction the energy
barriers are very small (0.06\,eV and 0.08\,eV); in fact they are only
marginally larger than that for the diffusion on the flat Al\,(111)
surface.

The exchange path at the \{100\}-faceted step is geometrically quite
different from that at the \{111\}-faceted step,\cite{Ex111S} as it has
no mirror symmetry perpendicular to the step.  We mapped out a
two-dimensional total-energy surface, varying the $x$- and
$y$-coordinates of the involved step atom (atom~2 in
Fig.~\ref{Diff100S}) on a 4$\times$4 grid while relaxing all the other
coordinates of the adsorbates and the two upper substrate layers. In
addition we calculated the energy for 4~points on the apparent diffusion
path. We then checked if all the atomic configurations were smoothly
connected along the diffusion path or if some atomic coordinates change
drastically between adjacent configurations. The results show that all
coordinates vary smoothly, which confirms that the described path is
physically relevant.

\section{Uniformity in surface self-diffusion on metals: energy barriers and
diffusion mechanisms}
\label{Unif}

It is interesting to compare our results for Al surface self-diffusion
to those for other fcc metals. The results in Table~\ref{Tab1} show that
the rule of thumb which says that diffusion barriers scale with the
cohesive energy is valid.  We also see, that the energy barriers for
surface self-diffusion are clearly lowest for the flattest surface,
namely (111).  All rougher surfaces have diffusion barriers about 5
times higher than the (111) surface of the same metal. The diffusion
barriers on these rougher surfaces vary by less than a factor of two
between different metals.

Important for a description of diffusion and its dependence on the local
geometry is the fact that in experiment (and in theory) the barriers for
diffusion in the channels of the (110) surface and along the
\{111\}-faceted steps of the (331) surface have nearly the same
barriers. The only counter example seems to be Ir, but there the
barriers for diffusion along the step represent only estimates. We
assume that the diffusion mechanism is the exchange (see
Fig.\,\ref{Kugeln}\,(d)\,) in both cases for all metals considered.

In agreement with our results for Al, the experiments show smaller
energy barriers for diffusion along the \{100\} step than along the
\{111\} step.  One might speculate that this difference of the barriers
is connected with the change in diffusion mechanism; we remind that for
Al we found that the diffusion along the \{100\} step proceeds by a
hopping mechanism while it proceeds via exchange along the \{111\} step.

Additional insights about the uniformity of diffusion at fcc\,(111)
surfaces can be obtained from the diffusion prefactors~$D_0$. Using
transition-state theory and semi-empirical calculations, $D_0$ was
evaluated for a series of metal surfaces.\cite{Liu91,Hansen91} In all
cases the prefactor of an exchange process was larger than that of the
normal hopping diffusion (see also Table~\ref{Tab2} for Al).
Furthermore, we see in Table~\ref{Tab2} that the experimentally
determined diffusion prefactors are larger in the cases where we expect
exchange diffusion. Thus, from the larger diffusion prefactors we
conclude that for Rh, Pt, and Ir there is hopping diffusion on the (111)
surface and along the \{100\}-faceted step, and exchange diffusion
otherwise (see Table~\ref{Tab2}).

\section{Atomic processes and growth of Al\,(111) at different temperatures}
\label{Atom}

The above discussed results for diffusion mechanisms, energy barriers,
and diffusion prefactors will now be used for an examination of the
temperature dependence of growth of Al\,(111). The mode of growth is
controlled by the interplay of the deposition rate and of the
temperature dependent processes of diffusion and defect creation and
annihilation. With the information of previous sections at hand we can
estimate the most important features of this temperature dependence. An
extended study with our results and employing a Monte-Carlo
technique\cite{Venables84,Kenny92,Smilauer93} to evaluate the relevant
rate equations, would be certainly superior.


Here we assume that the temperature dependence of the diffusion constant
is given by
  \begin{equation}
D= D_{0} \exp({-E_d/k_BT)}\;,
  \label{DifEq}
\end{equation}
 where $E_d$ is the energy barrier of the considered diffusion process
(see Table~\ref{D111} and \ref{D110}).  The pre-exponential factor
$D_{0}$ can be estimated from the adatom attempt frequency $\nu_a$ and
the distance between neighboring adsorption sites, $l$, as $D_0 = \nu_a
l^2/n$.  The factor $n$ equals 2 or 4, depending on the dimensionality
of the diffusion process.  $\nu_a$ can be calculated within
transition-state theory from the the total-energy
surface.\cite{Vineyard57,Gomer90} On the other hand, $D_{0}$ can also be
taken from experimental results for similar diffusion processes (see
Table~\ref{Tab2}).

Writing $D_0 = \nu_j l^2/n$ and defining the temperature $T_d$ at which a
certain diffusion mechanism
becomes active as that at which the adatom will jump at least once per second
($\nu_{j} = 1/$s), it follows that
\begin{equation}
   T_d = \frac{E_d}{k_B} / \ln \frac{n D_0}{\nu_j l^2}  \;. \label{iTD}
\end{equation}

In Table~\ref{Tab2} we give our choice for $D_0$, based on theoretical
results for the Al surface diffusion as well as on experimental data for Rh,
Pt, and Ir.
This choice is clearly not unambigous; one has to
consider, however, that an error of a factor of~10 in $D_{0}$ changes the
temperature $T_d$ by less than 10\,\%.
The results of the above procedure to determine onset temperatures for
diffusion are given in Table~\ref{EdTd}.

For temperatures below 320~K the desorption of adatoms from steps is
practically irrelevant (see, e.\,g., Fig.~\ref{Diff100S} and
Table~\ref{EdTd}).  Thus adatoms captured at a step edge will stay and the
island will grow. The processes important for growth at temperatures below
320~K are therefore the capture of Al adatoms at the steps and the diffusion
along the steps and their relative rates.

The following approximate
analysis of temperature ranges for different growth
modes assumes a deposition rate of $1/100$~ML/s.  The further
assumption is that for this deposition rate a jump rate of $\nu_j=1/$s
is large enough in order for a deposited Al adatom to diffuse to a
more stable site before other deposited adatoms meet the first one and
form a nucleus.  If the deposition rate is higher than $1/100$~ML/s
then the necessary jump rate to get the same growth mode has to be
increased proportionally.  This implies that the temperature has to
be increased logarithmically (see Eq.~\ref{iTD}).

The analysis implies:

\begin{itemize}
\item[--] Our calculations show that an Al dimer on Al\,(111) is bound by
0.58~eV and is therefore stable at temperatures below
$\simeq 250\,$K. If the mobility of the dimer is smaller than
that of the single adatom it will serve as a nucleus for the
growth of the next layer. In that case three-dimensional
growth would occur whenever two adatoms meet on the upper
terrace at a substantial rate.\cite{Bethge90}

\item[--] At temperatures below 25~K adatoms on the upper terrace are
hindered by the step-edge energy barrier to move down (see
Table~\ref{EdTd}).  We note that island edges are frayed at this
temperature (see next item), which may reduce the barrier,
and adatoms gain energy by adsorption and by approaching the upper step
edge, which might lead to some transient
mobility.\cite{Egelhoff89,Smilauer93} Furthermore we note that on small
and/or narrow islands the attempts to overcome the step-edge barrier are
more frequent which increases the chance for a success. We are therefore not
convinced that the three-dimensional growth mode really exists for clean
Al\,(111).

\item[--] For 25~K~$ < T < $~155~K the energy barriers of diffusion
parallel to both close-packed steps will prevent diffusion parallel to
the steps. The diffusion at flat parts of the surface is, however, still
easy and therefore the attraction of gas-like adatoms towards the steps
edges is still active.  As a consequence we expect that islands will be
formed in a way which may be described as a ``hit-and-stick''
mechanism. Thus, the edges cannot equilibrate and fractal-shaped islands
with a layer-by-layer growth mode should result.

\item[--] For $T > 155$~K the step edges will be straight, as diffusion
along the step is possible, and therefore the islands will be triangular
or hexagonal.  According to a simple model by Michely {\em et
al.}\cite{Michely93} the different diffusion properties for atoms at the
two kinds of step edges might become important for determination of the
detailed growth form of the island. The model says that the growth
perpendicular to the step with the lower adatom mobility will advance
fastest. If the growth speed of one step is faster by more than a factor
of two than that of the other step the faster one will eventually
disappear. As a consequence the growth shape of the island would be
triangular. For Al\,(111) our results imply that at low temperature the
diffusion parallel to the \{111\} step is hindered more than that
parallel to the \{100\} step because of the higher diffusion barrier.
However, because the diffusion mechanism is different at these two steps
the prefactor for the diffusion parallel to the \{100\} step is smaller
(see Table~\ref{Tab2}). As a consequence, at higher temperature the
diffusion coefficient for the movement parallel to the \{100\} step will
be smaller than that for the movement parallel to the \{111\}
step.\cite{crossover}

The additional assumption in Ref.\,\onlinecite{Michely93} is
that the growth speed of steps with larger diffusion barriers
along the step is higher. This is caused by the less efficient
atom transport along the step to corners of the growing
island. Atoms moving to the corners contribute to the growth
of the adjacent step.

We therefore predict that the adatom islands during growth at lower
temperature have shorter \{111\} edges and at higher temperature they
have shorter \{100\} edges. At a medium temperature the growth shape of
islands is hexagonal.

\item[--] There are two mechanisms for vacancy formation we considered
on the Al\,(111) surface.  The direct creation of vacancies on flat
Al\,(111) is expected to happen with a rate above 1/s per surface-atom
at temperatures above 730\,K (see Table~\ref{EdTd}). In the presence of
Al-adatoms the vacancy creation should already take place at 490\,K, due
to the then reduced barrier (see Table~\ref{EdTd}). Adatoms can be
provided either from deposition or by desorption from the step. The
relatively high energy barriers to create adatoms lets us assume, that
the adatom assisted vacancy formation will not be important before the
formation of vacancies on the flat Al\,(111) sets in at 730\,K.

  The second possibility to create vacancies is at steps.  At a
  \{100\}-faceted step they will be created with rates of $1/$s and
  step-atom at a temperature of 320\,K. At a \{111\}-faceted step this
  temperature is 380\,K (see Table~\ref{EdTd}).  These vacancies could
  migrate into the terrace and become ``normal'' surface vacancies.  The
  barrier for vacancy migration is 0.56\,eV on the flat Al\,(111), so it
  becomes active at 240\,K (see Tables~\ref{Tab1} and \ref{EdTd}). The
  onset temperature for vacancy generation from steps therefore should
  be 320\,K. Vacancy generation and diffusion into the terrace was also
  observed in STM experiments on Pt\,(111),\cite{Michely91} which again
  shows the similarity of atomic processes on Pt\,(111) and Al\,(111).
\end{itemize}

\section{Conclusion}

In conclusion, we have presented results of accurate
electronic structure and total-energy calculations which
reveal several phenomena directly relevant to the description
of self-diffusion at Al surfaces and to crystal growth.

The three low index surfaces of Al are quite different with regard to surface
self diffusion. The diffusion barriers for Al adatoms on Al\,(111)
[$E_d$\,=\,0.04\,eV] are much lower than on Al\,(100) and Al\,(110)
[$E_d$\,=\,0.33\,--\,0.62\,eV].  For Al\,(100) and Al\,(110) atomic exchange
mechanisms have lower barriers for surface self diffusion than ordinary
hopping.  Exchange diffusion was found even in the direction parallel to the
atomic rows on Al\,(110). The diffusion of surface vacancies was studied for
the Al\,(111) surface ($E_d$\,=\,0.56\,eV).

Our calculations predict that Al adatoms on Al\,(111) are
attracted towards the edge of close packed steps by a
long-range force which most likely originates from a
interaction of adatom and step induced surface states. Adatoms
close to the lower step edge are funneled towards the step.
The diffusion of an Al adatom from the upper to the lower
terrace proceeds via an exchange of the on-terrace adatom with
an in-step-edge substrate atom. The barrier for this
exchange process is rather low which leads to layer-by-layer
growth down to very low temperatures.  Corresponding results
have been observed in field-ion microscopy studies of W and Ir
adatoms on Ir\,(111) by Wang and Ehrlich.\cite{Wang91,Wang93}

On Al\,(111) the energy barrier for diffusion of an Al at-step
adatom parallel to the step is much bigger than that
perpendicular to the step in the descending direction.
In the temperature range where
this energy barrier becomes relevant, we expect fractal
growth.  The mechanism for diffusion along the two kinds of
steps is different.  Along the \{111\}-faceted steps we find
an atomic replacement mechanism similar to that for diffusion
parallel to the rows on Al\,(110), along the \{100\}-faceted
steps the hopping mechanism has the lowest energy barrier.
The differences in energy barrier and diffusion prefactor for
diffusion along the two kinds of steps can lead to temperature
dependent growth forms of islands.  Surface systems where
similar growth phenomena were observed experimentally are Pt
on Pt\,(111)\cite{Bott92,Michely93} and Au on
Ru\,(0001).\cite{Hwang91}

A comparison of our results for self diffusion on Al surfaces
and of other theoretical and experimental results on
transition metals indicate that atomic exchange processes in
surface self diffusion are more common than assumed
previously.

We predict that \{111\}-faceted steps on Al\,(111) are energetically favorable
compared to the \{100\}-faceted steps. A similar energy
difference was found for Pt\,(111) in experiment.\cite{Bott92}

Additionally to the energetics at Al surfaces we have discussed
the surface dipole moments induced by adatoms and steps on the Al\,(111)
surface and, related to that, the work function differences of the low
index surfaces of Al. Our results indicate that the commonly used model
based on Smoluchowski smoothing alone\cite{Zangwill88} has to be modified.


\appendix

\section{Al on Al\,(100)}
\label{Al100}

  We add here our results for the adsorption and diffusion of Al on
Al\,(100).  Our study repeats that of Feibelman on the same
system.\cite{Feibelman90} The main result of Feibelman's paper, which
is the favorable energy barrier for an exchange diffusion mechanism,
has been questioned recently.\cite{Liu91,Debiaggi92} We calculated the
adsorption energies at the three sites which are important for the
discussion of surface diffusion (see Fig.~\ref{Diff100View}). Our
results confirm that the exchange diffusion mechanism has a lower
barrier than the bridge diffusion mechanism. The agreement in adsorption
energy differences of our results with those of Feibelman is again as
excellent as in the other examples in this paper.  However, this
agreement with Feibelman's results is obtained only, if we use the same
slab thickness as he did, i.\,e., 5-layers (see Table~\ref{Diff100Dat}).
The agreement would be even better, if we had not relaxed all atoms but
only those which Feibelman had relaxed.  However, while the numerical
accuracy of both calculations agree, it is most interesting to note that
our calculations with 6 and 7 layer slabs show a significant change of
the energy of the exchange configuration. This change increases the
barrier for diffusion by nearly a factor of three (see
Table~\ref{Diff100Dat}).  A similar sensitivity of calculated energies
with slab thickness was not found for any other system and we do not
have an explanation for it. The energy of the exchange configuration
also proved to be especially sensitive to changes in the value of the
lattice constant and the {\bf k}-point sampling.

\newpage
\begin{table}
  \caption[]{Surface, step, adatom, and vacancy formation
energies for aluminum. The Al chemical potential is taken as
the cohesive energy, i.e.~4.15\,eV.\cite{cohesive} Thus, the
adatom is considered to be taken from a bulk or kink site, and
for the vacancy formation energies the removed atom is put
into such site.}\vspace*{5pt} \label{SurVacForm}
\begin{tabular}[t]{cdddd}
\squeezetable
system & \multicolumn{2}{c}{surface and step formation} & adatom formation&
 vacancy formation\\
 & (eV/atom) & (eV/\AA$^2$)& (eV)   & (eV) \\[1pt]
\hline
Al\,(111) & 0.48 & 0.070 & 1.05   & 0.67 \\
Al\,(100) & 0.56 & 0.071 & 0.38   & 0.65\\
Al\,(110) & 0.89 & 0.080 & 0.26   & 0.12\\[3pt]
$\langle 110\rangle/$\{111\}-step  &0.232 &  0.082 & 0.28 & 0.21 \\
$\langle 110\rangle/$\{100\}-step  &0.248 &  0.088 & 0.25 & 0.24
\end{tabular}
\end{table}

\begin{table}
\caption{\sloppy Total-energy difference $\Delta E$
per edge atom of two triangular
  islands on a 4-layer thick Al\,(111) slab in eV.
One island has only \{111\}- and the
  other one has only \{100\}-faceted steps. Four different island sizes are
  considered. Using a 5-layer substrate changes the results by $< 10\,\%$. The
  data in the rightmost column were obtained with the atoms of the islands
  relaxed. Relaxing more atoms does not change the energy differences
  significantly.} \vspace*{5pt}
\label{TriaDat}
\begin{tabular}{ddddd}
\squeezetable
\# atoms &  \# edge atoms  & surf.\ cell &  $\Delta E^{\rm unrelaxed}$ &
$\Delta E^{\rm relaxed}$\\[2pt] \hline
6          &       6  & $6\times5$ &    0.025 &  0.029 \\ 
10         &       9  & $6\times5$ &    0.019 &  0.021 \\ 
15         &      12  & $6\times6$ &    0.017 &  0.018 \\ 
21         &      15  & $8\times7$ &    0.017 &  0.018 \\ 
\end{tabular}
\end{table}

\begin{table}
\caption[]{Induced dipole moment $\mu$ of Al adsorbates on fcc and hcp sites
of Al\,(111) at 1/16\,ML coverage and of a step atom in \{111\}- and
\{100\}-faceted steps on Al\,(111) in debye.\cite{debye} Positive
$\mu$ means that the negative end of the dipole points into the vacuum.
Results are given for the substrate atoms unrelaxed and
relaxed. The numerical accuracy of the given values is
$\pm0.01$\,debye. The values are averages for slabs of
5 to 7 layers thickness.
}\vspace*{5pt}
\label{TabDip}
\begin{tabular}{ldd}
system & $\mu^{\rm{unrelaxed}}$ & $\mu^{\rm{relaxed}}$\\[2pt]\hline
fcc-site adatom  & 0.13 & 0.30\\
hcp-site adatom  & 0.06 & 0.24\\ [2pt]
\{111\}-faceted step & --0.01 & --0.01\\
\{100\}-faceted step & 0.045 & 0.045\\
\end{tabular}
\end{table}

\begin{table}
\caption[]{Total energies for an isolated Al
  adatom on Al\,(111) at fcc, bridge, hcp, and top sites and
on the fcc or hcp site directly at the upper side of the
\{111\}- and the \{100\}-faceted step.  The energy zero is the
energy of a free aluminum atom.~\cite{cohesive}
 For the adsorption on the flat
Al\,(111) surface also the adsorbate height is given with
respect to the center of the top substrate
layer.}\vspace*{5pt}
\begin{tabular}{lccdd}
site  & coordination & $E\;$(eV) & $h\;$(\AA)\\[2pt]\hline
fcc  & 3 & $-$3.06 & 2.11\\
bridge & 2 & $-$3.06 & 2.09\\
hcp & 3 & $-$3.10 & 2.08\\
top & 1 & $-$2.57 & 2.12\\[3pt]
fcc on $\langle 110\rangle/$\{111\}-step & 3 &  $-$3.18 & ---\\
hcp on $\langle 110\rangle/$\{100\}-step & 3 & $-$3.18  & ---
\end{tabular}
\label{D111}
\end{table}

\begin{table}
\caption[]{
  Comparison of calculated energy barriers (in eV) for surface self-diffusion
on Al
  with those by embedded-atom calculations of Liu et al.\cite{Daw84} for Al
  (two potentials were used there; both results deviate considerably from
  ours) and with experimentally determined barriers on other metal surfaces.
  The experimental results were determined using field ion microscopy. Values
  in brackets are believed to be less accurate, as they were obtained
  with an assumed value for the diffusion prefactor $D_0$. The symbols
$\parallel$ and $\perp$ indicate a diffusion direction parallel or
perpendicular to the channels of the (110) surface or to the step edge
respectively.}\vspace*{5pt}
 \label{Tab1}
{\centering
\begin{tabular}{l|d|dd|dddd}
surface & Al (this\ work) & \multicolumn{2}{c|}{Al\cite{Liu91}} &
Ni\cite{Liu91,Tung80}& Rh\cite{Ayrault74}& Pt\cite{Kellogg91,Basset78}&
Ir\cite{Wang89,Tsong91,Tsong90,Wang90} \\[2pt] \hline
(111)
   &0.04 & 0.054& 0.074&   ---      &0.16 &(0.12)&0.27\\
vacancy at Al\,(111) & 0.56 & ---& ---& ---& ---& --- \\[2pt]
(100)
   &0.35 & 0.69& 0.25 &     ---     &0.63 &0.47 &0.84
\\[2pt](110)$\parallel$
   &0.33 & 0.26& ---& (0.45)&0.60 &0.84 &0.80\\
(110)$\perp$
   &0.62 & 0.30& 0.15             &(0.45)&   ---     &0.78 &0.71
\\[2pt]$\langle 110\rangle/$\{111\}-step$\parallel$ or (332)$\parallel$
   &0.42 & 0.27& 0.24&  (0.45)&0.64 &0.84 &(1.05)\\
$\langle 110\rangle/$\{100\}-step$\parallel$ or (644)$\parallel$
   &0.32 & 0.20& 0.24&  (0.37)&0.54 &0.69 &(0.96)\\[2pt]
\hline
cohesive energy\cite{Kittel}&
  \multicolumn{3}{c|}{3.39} & 4.44&5.75 &5.84 &6.94
\end{tabular}
}
\end{table}

\begin{table}
 \caption[]{Total energies $E$ for Al adatoms with respect to that of a
free Al atom and energy barriers $\Delta E$ (both in eV) for sites with
similar local geometry on the Al\,(110), the Al\,(331) surface and at
steps on Al\,(111) (compare Fig.~\ref{Kugeln}).~\cite{cohesive} The
results for the (331) surface may be compared to those for the $\langle
110 \rangle$/\{111\} step. For the adsorption on Al\,(110) we give also
the height $h$~(in \AA) above the relaxed, flat surface.  As explained
in the text, the exact barriers for exchange diffusion parallel to the
step edge might be 0.04\,eV higher than given in the
table.}\vspace*{5pt}
  \begin{tabular}{lddddd}
  \squeezetable & \multicolumn{2}{c}{Al\,(110)} &
    Al\,(331)\tablenote{Calculations by Feibelman;\cite{Feibelman92} for
    technical differences from our calculations see text and
    reference~\onlinecite{Feibelman92}}
 &  $\langle 110\rangle/$\{111\}-step & $\langle 110\rangle/$\{100\}-step\\
& $E$ & $h$ & $E$ &
$E$ & $E$\\[1pt]\tableline
(a) fivefold site & $-$3.89 & 1.33 & $-$3.68 & $-$3.87 & $-$3.90\\[4pt]
diffusion & $\Delta E$ & & $\Delta E$ &
$\Delta E$ & $\Delta E$ \\[1pt] \tableline
(b) long bridge
& --0.60 & 1.58 & --0.57 & --0.48 & --0.32 \\
(c) short bridge
& --1.06 & 2.16 & --1.21 & --1.03 & --1.15 \\[3pt]
(d) exchange $\parallel$ 
& --0.33 & 1.27 & --- & --0.39 & --0.44\\
(e) exchange $\perp$ & --0.62 & 0.80 & --- &
\multicolumn{1}{c}{--0.76\,(--0.06)\tablenote{In brackets we give the barrier
    for the descending diffusion}}  & --0.80\\
\end{tabular}
\label{D110}
\end{table}

\begin{table}
  \caption[]{Diffusion prefactors $D_0$ (in cm$^2$/s) from theory for Al
(mean of the two values given in Ref.~\onlinecite{Liu91}) and from
experiment for Rh, Pt, and Ir surfaces. Values in brackets are
considered to be less reliable. The (331) surface has \{111\}-faceted
steps and the (311) surface has \{100\}-faceted steps. The column
``mechanism'' contains our assumptions about the mechanism of diffusion
for every row, and the right column gives the diffusion prefactors which
will be used in the temperature dependence of growth of
Al\,(111).\label{Tab2}}\vspace*{5pt} \centering
  \begin{tabular}{lc|cc|ccc|c}
surface & mechanism & \multicolumn{2}{c}{Al\cite{Liu91}}&
Rh\cite{Ayrault74}& Pt\cite{Kellogg91,Basset78}&
Ir\cite{Wang89,Tsong91,Tsong90,Wang90}& our choice\\ [2pt]
(111) & hopping
   &$9\times10^{-4}$ &$1.6\times10^{-3}$ &$2\times10^{-4}$ &$ (3\times10^{-4})$
&$9\times10^{-5}$ & $2\times10^{-4}$\\
(100) & exchange
   & --- &$4\times10^{-2}$ &$(1\times10^{-3})$ & $1.3\times10^{-3}$
   &$6\times10^{-2}$ & $8\times10^{-3}$\\ [3pt]
(110)$\parallel$ & exchange
   & --- & ---  
&$3\times10^{-1}$ &$8\times10^{-3}$ &$6\times10^{-2}$  &
$1\times10^{-2}$ \\
(110)$\perp$ & exchange
   &$6\times10^{-2}$ &$2.4\times10^{-2}$ & --- & $1\times10^{-3}$
&$4\times10^{-3}$ & $2\times10^{-3}$\\[3pt]
(331)$\parallel$ & exchange
   & --- & --- 
&$1\times10^{-2}$ &$4\times10^{-4}$ & --- & $1\times10^{-2}$\\
(311)$\parallel$ & hopping
   &$2\times10^{-3}$ &$6.7\times10^{-3}$ &$2\times10^{-3}$ &
   $(1\times10^{-6})$ & --- & $5\times10^{-4}$
\end{tabular}
\end{table}

\begin{table}
\caption[]{
  Energy barriers $E_d$ (in eV) for different self-diffusion and
vacancy-formation processes on Al surfaces.  From these barriers and
from estimates of the pre-exponential $D_0$ in Eq.\,\ref{DifEq} (see
Table~\ref{Tab2}) we calculate the temperatures $T_d$ at which these
processes happen at a rate of $1/s$ per atom (see Eq.\,\ref{iTD}).
Exchange processes are indicated.\\
Note that the thermodynamical vacancy
formation energies as given in Table~\ref{SurVacForm}
are lower than the vacancy formation barriers.
\label{EdTd}}\vspace*{5pt}
{\centering
\begin{tabular}{ldd}
{\em adatom diffusion} &  $E_d$ (eV)& $T_d$ (K)\\[3pt]
flat Al\,(111)
   &0.04 & 17\,$\pm$10 \\
flat Al\,(100)  \hfill[exch.] &0.35 & 135\,$\pm$23\\[3pt]
Al\,(110)\,$\parallel$ to rows \hfill[exch.]
   &0.33 &  130\,$\pm$23 \\
(110)\,$\perp$ to rows \hfill[exch.]
   &0.62 & 245\,$\pm$34 \\[3pt]
$\langle 110\rangle/$(111) step\,$\parallel$ \hfill[exch.]&0.42 &
155\,$\pm$25\\
$\langle 110\rangle/$(100) step\,$\parallel$ & 0.32 & 135\,$\pm$23 \\[3pt]
$\langle 110\rangle/$(111) step\,$\perp$ descending  \hfill[exch.]
& 0.06 & 25\,$\pm$12  \\
$\langle 110\rangle/$(100) step\,$\perp$ descending \hfill[exch.]
& 0.08 & 33\,$\pm$13  \\[3pt]
{\em other processes on Al\,(111)} \\[3pt]
vacancy diffusion on Al\,(111) & 0.56 & 240\,$\pm$35 \\
adatom desorption  from step &
\multicolumn{1}{c}{$\simeq 0.8$} & \multicolumn{1}{c}{$\simeq 320$}\\
vacancy-formation in $\langle 110\rangle/$\{100\}
step\tablenote{Estimated energy barriers, assuming that the
  transition state is similar to that for bridge diffusion along the
  step (see Table\,\ref{D110}).}
& \multicolumn{1}{c}{$\simeq 0.8$} & \multicolumn{1}{c}{$\simeq 320$} \\
vacancy-formation in $\langle 110\rangle/$\{111\} step$^{\rm a}$
& \multicolumn{1}{c}{$\simeq 0.95$} & \multicolumn{1}{c}{$\simeq 380$} \\
vacancy-formation on flat surface\tablenote{The assumed transition state for
  the higher of the two values is
  that for bridge diffusion across the step (see Table\,\ref{D110}). The
  lower value corresponds to vacancy-formation in the presence of another
  Al adsorbate.} &
\multicolumn{1}{c}{$1.2-1.8$} & \multicolumn{1}{c}{$490 - 730$}
\end{tabular}
}
\end{table}

\begin{table}
 \caption[]{Adsorption energies (in eV) and heights (in
bohr) for Al adsorbed on
Al\,(100) at 1/16\,ML coverage. The energy zero is the energy of an
isolated, free Al atom. The considered configurations are pictured in
Fig.~\ref{Diff100View}. Results for slabs of different thickness are
compared with those obtained by Feibelman\cite{Feibelman90} who used a
5-layer slab. He used the experimental lattice constant of 7.66\,bohr
and allowed only the adsorbate and its substrate neighbors to relax (see
also Ref.~\onlinecite{Feibelman92}). On the other hand, we use the
theoretical lattice constant of 7.56\,bohr, one special {\bf k}-point in
the surface Brillouin zone, and we allow the adsorbate and the upper two
layers to relax.  For the results labeled as ``average'' additional
calculations with 4~{\bf k}-points and with an additional layer relaxed
were considered as well.  Energies are in eV, the adsorbate heights $h$
are in \AA{} relative to the relaxed clean surface.}
    \leavevmode
    \begin{tabular}{lcddd}
 & configuration & $E$ & $\Delta E^{\rm 4-fold}$ &  $h$\\[1pt]
5-layer (Ref.~\onlinecite{Feibelman90}) & 4-fold & $-$2.93 & ---      & 1.72\\
                       & bridge & $-$2.28 & 0.65   & 2.20\\
                       & exchange & $-$2.73 & 0.20 & 0.90\\[1pt]\hline
5-layer  & 4-fold & $-$3.68 & ---& 1.58\\
& bridge & $-$3.05& 0.63 & 1.91 \\
& exchange & $-$3.55 & 0.13 & 0.74 \\[1pt]\hline
6-layer & 4-fold & $-$3.75 & ---& 1.70 \\
& bridge & $-$3.12 & 0.63 & 2.11 \\
& exchange & $-$3.37 & 0.38 & 0.91 \\[1pt]\hline
7-layer & 4-fold  & $-$3.75 & --- & 1.73 \\
& bridge & $-$3.07 & 0.69 & 2.12 \\
& exchange & $-$3.35 & 0.40 & 0.98\\[1pt]\hline
``average'' & 4-fold  & $-$3.77 & --- & 1.69\\
& bridge & $-$3.12 & 0.68 & 2.09 \\
& exchange & $-$3.42 & 0.35 & 0.90\\[1pt]
    \end{tabular}
  \label{Diff100Dat}
\end{table}
\clearpage
\begin{figure}
\caption{Side view (upper panel) and top view (lower panel) of an
  orthorhombic supercell, containing a four-atom wide terrace stripe,
   oriented
  in $\langle 110 \rangle$ direction, on an Al\,(111) substrate. Note the
  different step types at the left terrace edge (a \{100\} micro-facet) and
  right terrace edge (a \{111\} micro-facet). The two step types are labeled
  as $\langle 110\rangle /\{100\}$ and $\langle 110\rangle /\{111\}$. The three
  sites important for diffusion on flat Al\,(111) are also indicated.}
\label{StepScheme}
\end{figure}

\begin{figure}
\caption[]{Top and side view of the fcc\,(332) surface. The
  (332) surface has \{111\}-faceted steps and the
 number of atomic rows within the (111)
oriented terraces is six.}\label{Systeme1}
\end{figure}
\begin{figure}
\caption[]{Top and side view of the fcc\,(433) surface. The
  (433) surface has \{100\}-faceted steps and the
 number of atomic rows within the (111)
oriented terraces is seven.}\label{Systeme2}
\end{figure}

\begin{figure}
\caption[]{
View at islands on a fcc\,(111) surface. The two differently oriented
triangular islands have only one kind of step; the hexagonal island has
both sorts of step.}
\label{Triangle}
\end{figure}

\begin{figure}
\caption{Total energy along the diffusion path of an Al adatom crossing
  a \{111\}-faceted step on Al\,(332) (see also
  Fig.\,\protect{\ref{Systeme1}}).  The upper curve is calculated for the
  ``normal'' hopping diffusion and the lower one for the exchange
  process.  The generalized coordinate is $Q= X_{1} + X_{2}$ where
  $X_{1}$ and $X_{2}$ are the $x$-coordinates of the adatom labeled as
  No.~$1$ in the atomic-structure plot and $X_{2}$ is the position of a
  step-edge atom labeled as No.~$2$.  The $x$-axis is parallel to the
  surface and perpendicular to the step orientation.  For the
  undistorted step $X_{2}=0$.  All other coordinates of the diffusing
  atom and of the two top substrate layers are optimized for each
  position $Q$.}
\label{Diff111S} \end{figure}

\begin{figure}
\caption[]{
  \small Upper panel: total energy along the diffusion path on an
Al\,(433) surface for the generalized coordinate $Q = X_1 + X_2$
belonging to the two atoms labeled~1 and~2, which are involved in the
exchange process for the across step diffusion (see also
Fig.\,\ref{Diff111S}).\\ Medium panel: Top view of the Al adatom
situated on top of the
\{100\}-faceted step. The rectangle gives the range of
$x$-$y$-coordinates at which atom~2 was set for finding the
lowest-energy path (see also Fig.\,\ref{Systeme2}).\\
  Lower panel: Contour plot of the total energy of the system with the
$x$-$y$ coordinate of atom No.~2 fixed at positions in a regular
$4\times4$ mesh in the rectangle in the medium panel (contour spacing
0.04\,eV). All other
coordinates of the adsorbates and the  two top layers were relaxed. The
dashed line connects equivalent points in the two figures, the dashed
quarter circles indicate the in-step and the at-step position of atom
No.~2.}\label{Diff100S}
\end{figure}

\begin{figure}
 \caption[]{Important adatom geometries on the (110) surface (top) and
at the \{111\}-faceted step on a fcc(111) surface (bottom). The energies
of these geometries are given in Table~\ref{D110}.} \label{Kugeln}
\end{figure}

\begin{figure}
 \caption{View at the 3 adsorption geometries considered for the Al
self-diffusion on Al\,(100).} \label{Diff100View}
 \end{figure}

\vspace{1cm}
\begin{references}
\bibitem[*]{Albu} present address: Sandia National Laboratories,
Division 1114, Albuquerque, NM 87\,185-0344

\bibitem{Ayrault74} G. Ayrault and G. Ehrlich, J. Chem. Phys. {\bf 60}, 281
(1974).

\bibitem{Wang89} S.\,C. Wang and G. Ehrlich, Phys.~Rev.~Lett. {\bf 62}, 2297
(1989).

\bibitem{Kellogg91} G.\,L. Kellogg, Surf.~Sci.~{\bf 246}, 31 (1991).

\bibitem{Tsong91} T.\,T.~Tsong and C.-L.~Chen, Phys.\ Rev.~B~{\bf 43}, 2007
(1991).

\bibitem{Tsong90} T.\,T. Tsong, {\em Atom-probe field ion
microscopy\/}, Cambridge University Press (1990).

\bibitem{Bethge90} H. Bethge, p.~125 in {\em Kinetics of Ordering and Growth at
Surfaces}, ed. by M. Lagally, Plenum Press (New York, 1990).

\bibitem{Klaua85} M. Klaua, H. Bethge, Ultramicroscopy {\bf 17}, 73 (1985)

\bibitem{Kunkel90} R.~Kunkel, B.~Poelsema, L.K.~Verheij, and G.~Comsa, Phys.\
Rev.\ Lett.~{\bf 65}, 733 (1990).

\bibitem{Bott92} M.~Bott, T.~Michely, and G.~Comsa, Surf.\ Sci.~{\bf 272}, 161
(1992).

\bibitem{Michely93} T. Michely, M. Hohage, M. Bott, and G. Comsa, Phys.
Rev. Lett. {\bf 70}, 3943 (1993).  

\bibitem{Hwang91} R.Q. Hwang et al., Phys. Rev. Lett. {\bf 67}, 3279 (1991).

\bibitem{Wang90} S.\,C. Wang and G. Ehrlich, Surf.~Sci.~{\bf 239}, 301 (1990).

\bibitem{Egelhoff89} W.\,F. Egelhoff Jr. and I. Jacob, Phys.\ Rev.\ Lett.~{\bf
  62}, 921 (1989).

\bibitem{Hinch93} B.\,J.~Hinch, R.\,B.~Doak, and L.\,H.~Dubois, Surf.\
Sci.~{\bf 286}, 261 (1993). 

\bibitem{Oppo93} S. Oppo, V. Fiorentini, and M. Scheffler,
Phys.~Rev.~Lett.~{\bf 71}, 2437 (1993).

\bibitem{Vineyard57} G. Vineyard, J. Phys. Chem. Solids {\bf 3}, 121 (1957).

\bibitem{Gomer90} R. Gomer, Rep.~Prog.~Phys {\bf 53} 917-1002 (1990).

\bibitem{Venables84} J.\,A.~Venables, G.\,D.\,T.~Spiller, and M.~Hanb\"ucken,
Rep.\,Prog.\,Phys.~{\bf 47} (1984).


\bibitem{hydro} J. Villain, J. Phys. I {\bf 1} 19 (1991);
Z.-W. Lai and S. Das Sarma, Phys.\,Rev.\,Lett.~{\bf 66}, 2348 (1991);
Hong Yan, Phys.\,Rev.\.Lett.\, {\bf 68}, 3048 (1992).

\bibitem{Kenny92} S. Kenny, M.\,R. Wilby,  A.\,K.
Myers-Beaghton, and D.\,D. Vvedensky, Phys.~Rev.~B {\bf 46}, 10\,345 (1992).

\bibitem{Smilauer93} P.~\v{S}milauer, M.\,R. Wilby, and D.\,D.  Vvedensky,
Phys.\ Rev.\ B {\bf 47}, 4119 (1993).


\bibitem{Hammonds92} K.D. Hammonds and R.M. Lynden-Bell, Surf.~Sci.~{\bf 278},
437 (1992).

\bibitem{Liu91} C.\,L.\ Liu, J.\,M.\ Cohen, J.\,B.\ Adams, and A.\,F.  Voter,
Surf.\ Sci.~{\bf 253}, 334 (1991).

\bibitem{Liu92} C.-L. Liu and J.B. Adams, Surf.  Sci. {\bf 265}, 262 (1992).

\bibitem{Nelson93} R.\,C. Nelson, T.\,L. Einstein, S.\,V.  Khare, and P.\,J.
Rous, Surf.\ Sci.\ {\bf 295}, 462 (1993).

\bibitem{Hansen91} L.\,B. Hansen, P. Stoltze, K.\,W. Jacobsen, and J.\,K.
N\o{}rskov, Phys.~Rev.~B {\bf 44}, 6523 (1991); L.\,B. Hansen, P. Stoltze,
K.\,W. Jacobsen, and J.\,K. N\o{}rskov, Surf.~Sci.~{\bf 289}, 68 (1993).

\bibitem{Smoluchowski41} R.\,Smoluchowski, Phys.\,Rev.\,{\bf 60}, 661 (1941).

\bibitem{Zangwill88} A. Zangwill, {\em Physics at Surfaces\/}, University
Press, Cambridge (1988).

\bibitem{Pickett89} W.\,E. Pickett, Comp.\,Phys.\,Rep.~{\bf 9}, 117 (1989).

\bibitem{Scheffler92} Some of the calculations presented here were reported
previously in M.~Scheffler,
J.~Neugebauer, and R. Stumpf, J.~Phys.: Cond.\ Matter {\bf 5}, A91 (1993), in
R. Stumpf and M. Scheffler, Phys.\,Rev.\,Lett.\,{\bf 72}, 254 (1994), and in
R. Stumpf and M. Scheffler, Surf.~Sci.~{\bf 307-309}, 501 (1994).


\bibitem{Lang72} B. Lang, R.\,W. Joyner, and G.\,A. Somorjai, Surf.~Sci.~{\bf
  30}, 440 (1972); M.\,A. van Hove and G.\,A. Somorjai, Surf.~Sci.~{\bf 92},
489 (1980); D.\,R.~Eisner and T.\,L.~Einstein,  Surf.~Sci.~{\bf 286},
L559 (1993).

\bibitem{Michely91} T. Michely and G. Comsa, Surf.~Sci.~{\bf 256}, 217 (1991);
T. Michely, T. Land, U. Littmark, and G. Comsa, Surf.~Sci.~{\bf 272}, 204
(1992).  

\bibitem{Chen93} C.-L. Chen and T.~T. Tsong, Phys.\,Rev.\,B~{\bf 47}, 15\,852
(1993).

\bibitem{Wang91} S.C. Wang and G. Ehrlich, Phys. Rev. Lett. {\bf 67},
2509 (1991).  

\bibitem{Besocke77} K. Besocke, B. Krahl-Urban, and H. Wagner, Surf. Sci. {\bf
  68}, 39 (1977).

\bibitem{Stumpf94b} R. Stumpf and M. Scheffler, Computer
Physics Communications,  {\bf 79}, 447,  Cat. No. ACTF (1994).

\bibitem{Ceperley80} D.\,M.~Ceperley and B.\,J.~Alder, Phys.\ Rev.\ Lett.~{\bf
  45} (1980) as parameterized by J.\,P.~Perdew and A.~Zunger, Phys.\
Rev.~B~{\bf 23 }, 5048 (1981).

\bibitem{Kohn65} W. Kohn and L. J. Sham, Phys. Rev.  {\bf 140}, A1133 (1965).

\bibitem{Car85} R. Car and M. Parrinello, Phys. Rev. Lett. {\bf 55}, 2471
(1985).

\bibitem{Williams87} A. Williams and J. Soler, Bull.\ Am.\ Phys.\ Soc.~{\bf
  32}, 562 (1987).

\bibitem{Stumpf90} R. Stumpf, X. Gonze, and M. Scheffler,
Research report of the Fritz-Haber-Institut, April (1990), and
 X.~Gonze, R.~Stumpf, and M.~Scheffler, Phys.\ Rev.~B~{\bf
  44} 8503 (1991).

\bibitem{Monkhorst76} H.\,J.~Monkhorst and J.\,D.~Pack, Phys.\ Rev.~B~{\bf
  13}, 5188 (1976).


\bibitem{Gillan89} M.\,J. Gillan, J. Phys.\ Cond.\ Matt.: {\bf 1}, 689 (1989).

\bibitem{Neugebauer92} J. Neugebauer and M. Scheffler, Phys.~Rev.~B~{\bf 46},
16\,067 (1992).

\bibitem{Vita91} A. de Vita and M.\,J. Gillan, J.~Phys.~C:
Cond.~Matt.~{\bf 3}, 6225 (1991). 

\bibitem{Kresse93} G. Kresse, J. Haffner, Phys.~Rev.~B~{\bf 48},
13\,115 (1993).

\bibitem{Kittel} C.\,Kittel, {\em Introduction into Solid State Physics}, 6th
Edition, Wiley (1986).

\bibitem{MP} There exists a more elegant way to smear the
occupation numbers around the Fermi energy introduced in
M. Methfessel and A.\,T. Paxton, Phys.\ Rev.~B~{\bf
  40}, 3616 (1989). There the total energy is minimized so
that the energies and forces do not have to be corrected. We
therefore used this method in later studies.

\bibitem{Pederson91} M.\,R. Pederson and K.\,A. Jackson, Phys.\,Rev.\,B~{\bf
  43}, 7312 (1991).

\bibitem{Arias92} K.-M. Ho,
J. Ihm, and J.\,D. Joannopoulos, Phys.\,Rev.~B~{\bf 25}, 4260 (1982), and
T.\,A. Arias, M.\,C. Payne, and J.\,D. Joannopoulos, Phys.\,Rev.~B {\bf
45}, 1538 (1992).

\bibitem{Finnis90} M.\,W. Finnis, J.~Phys.: Cond.~Matt. {\bf 2}, 331 (1990).

\bibitem{conjgrad} The gain in speed by using more sophisticated update
techniques like the conjugate gradient approach\cite{Gillan89} would be
small for large metallic systems, according to our experience.

\bibitem{Nelson92} J.\,S.~Nelson and P.\,J.~Feibelman, Phys.\,Rev.\,Lett.~{\bf
  68}, 2188 (1992).

\bibitem{Feibelman92} P.\,J. Feibelman, Phys.~Rev.~Lett.~{\bf 69}, 1568
(1992).  The adsorption energies given in this paper are not directly
comparable to ours (see Ref.~\onlinecite{cohesive}) because of the
reference energy of the isolated atom. Apparently there is an
inconsistency of this reference energy caused by Feibelman's
Greens-function technique; energy differences are comparable, however,
and for these the main (but small) differences to our results arise
because Feibelman did not include the adsorbate-induced relaxation of
the Al\,(331) substrate.

\bibitem{Wang93} S.C. Wang and G. Ehrlich, Phys. Rev. Lett. {\bf 70}, 41
(1993).

\bibitem{cohesive} Cohesive and adsorption energies are given with
respect to the energy of an isolated Al atom calculated in a
large cell with the same 8\,Ry cut-off.  Adding the
spin-polarization energy of 0.15\,eV gives the cohesive energy
as 4.15\,eV which is 0.75\,eV higher than the experimental
value. This overbinding is a problem common to converged
DFT-LDA calculations. It is widely accepted that adsorption
energy {\em differences} for different sites are practically
unaffected by this problem. An Al atom adsorbed at a kink site
gains exactly the cohesive energy.




\bibitem{embedding} The embedding charge density is the central quantity
in embedded atom and effective medium theory.\cite{Daw84,Jacobsen87} The
embedding density is closely connected to the coordination number of the
adsorbate, so that coordination and embedding may in fact be regarded as
different names for the same thing.


\bibitem{Daw84} M.\,S.~Daw and M.\,I.~Baskes, Phys.~Rev.~B, {\bf 29}, 12
(1984).

\bibitem{Jacobsen87} K.\,W.~Jacobsen, J.\,K.~N\o{}rskov, and M.\,J.~Puska,
Phys.Rev.~B {\bf 35}, 7423 (1987).

\bibitem{Bondcut} ``Simple bond-cutting'' models assume that the energy
per atom varies linearly with the atom's coordination number.  An
improved version which takes the bond saturation into account makes this
approach very similar to the effective-medium and embedded-atom methods
(see for example I.J Robertson et al., Europhys.~Lett.~{\bf 15}, 301
(1991), Phys.~Rev.~Lett.~{\bf 70}, 1944 (1993), and M.~Methfessel et
al., Appl.~Phys.~A {\bf 55}, 442 (1992)\,).

\bibitem{debye} A dipole moment of 1 debye equals $2.4 e$/\AA{} were
$e$ is the electronic charge.


\bibitem{Grepstad76} J.\,K.\ Grepstad, P.\,O.\ Gartland, and
B.\,J.\ Slasvold, Surf.\,Sci. {\bf 57}, 348 (1976).

\bibitem{averlay} The values for the work functions were
determined by averaging values for slabs of thickness 5 to 7
layers for Al(111) and 8 and 9 layers for Al(110).


\bibitem{Hoelzl79} J. H\"olzl and F.\,K. Schulte, in {\em Springer
Tracts in Modern Physics\/} {\bf 85}, p.~1-100, Springer, Berlin (1979).

\bibitem{Ishida92} H. Ishida and A. Liebsch, Phys.\,Rev.\,B {\bf 46}, 7153
(1992).

\bibitem{Ari94} A.\,P. Seitsonen and M. Scheffler, in preparation.

\bibitem{Stampfl94} C.~Stampfl, M. Scheffler, H. Over, J. Burchhardt, M.
Nielsen, D. L. Adams, and W. Moritz, Phys. Rev. B, accepted; Phys. Rev.
Lett.

\bibitem{Hammer92} B. Hammer, K.\,W. Jacobsen, V. Milman, and M.\,C.
Payne, J. Phys.: Cond.~Matt.~{\bf 4}, 10\,453 (1992).


\bibitem{Tung80} R.\,T. Tung and W.\,R. Graham, Surf.~Sci.~{\bf 97}, 73
(1980).

\bibitem{Basset78} D.W. Basset and P.R. Webber, Surf. Sci {\bf 70}, 520
(1978).

\bibitem{Crommie93} M.\,F. Crommie, C.\,P. Lutz, and D.\,Eigler, Nature~{\bf
  363}, 524 (1993).

\bibitem{Hasegawa93} Y. Hasegawa and P. Avouris, to be published (1993).

\bibitem{Ex111S} There could be a nonsymmetric exchange path at the
\{111\}-faceted step also. We did not calculate that path as the
symmetric one has already such a low barrier which is at the limit of
the accuracy of our calculations. Our conclusions therefore would not be
affected by an additional diffusion process. No energetically favorable
symmetrical exchange path exists for the \{100\}-faceted step, as there
the involved step atom would have to go over a top site.

\bibitem{Feibelman90} P.\,J.\,Feibelman, Phys.\,Rev.\,Lett.~{\bf 65}, 729
(1990). 

\bibitem{Debiaggi92} S. Debiaggi and A. Caro, J. Phys.: Cond.~Matt. {\bf 4},
3905 (1992).










\bibitem{crossover} We can estimate the temperature $T^{\rm cross}$
at which diffusion
along the two kinds of steps will proceed at the same rate.
Transforming Eq.\,\ref{DifEq} to
  \begin{displaymath}
  T^{\rm cross} = k \frac{E^{\{111\}\,\parallel}_d
-E^{\{110\}\,\parallel}_d} {\ln
    D^{\{100\}\,\parallel}_0 - \ln D^{\{111\}\,\parallel}_0}
  \end{displaymath}
and taking the values for $D_0$ and $E_d$ for the
$\langle 110\rangle/$\{111\}- and $\langle 110\rangle/$\{100\} step from
Tables\,\ref{Tab2} and~\ref{EdTd}, we get $T^{\rm cross} \simeq 400$\,K.
We expect that this value is rather inaccurate however.




\end{references}
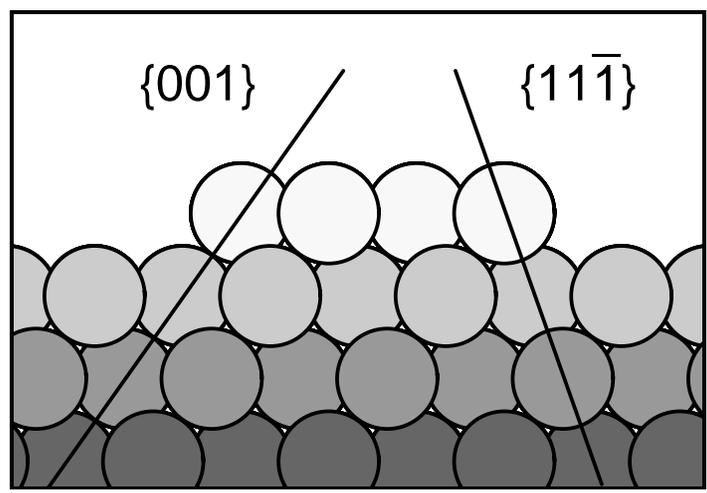
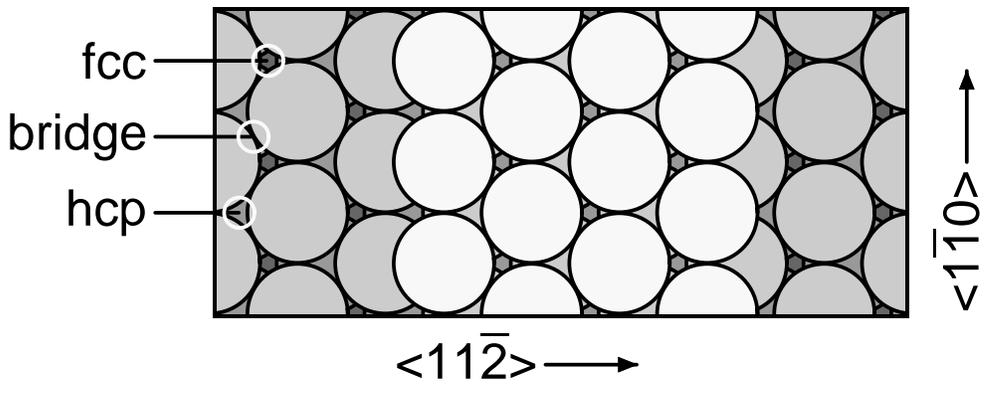
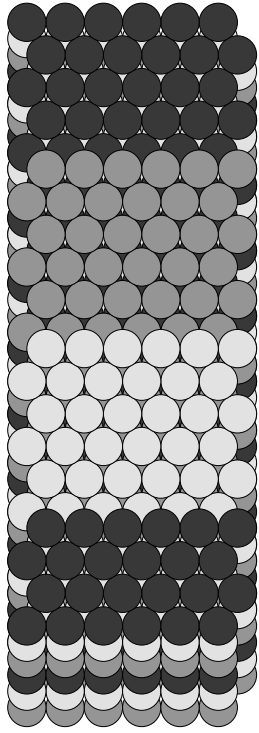
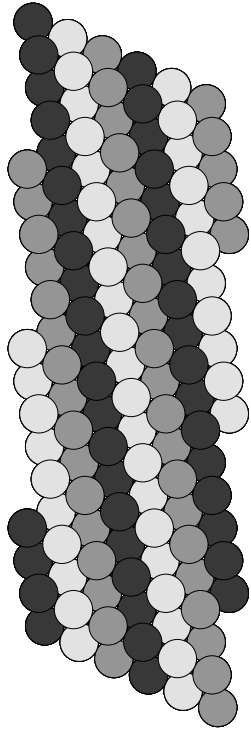
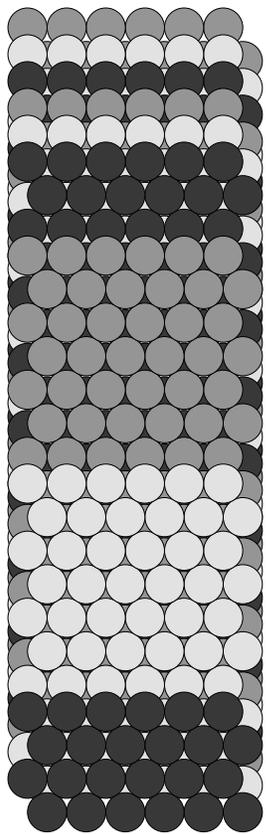
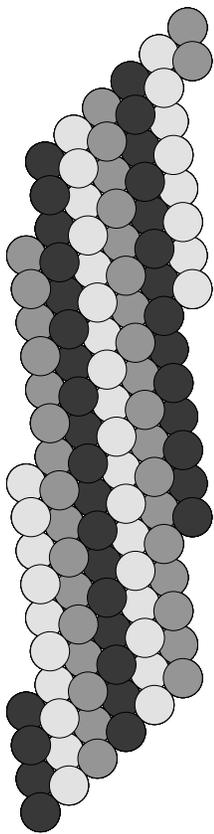
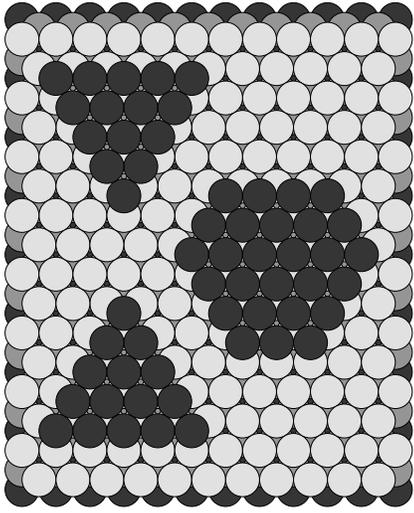
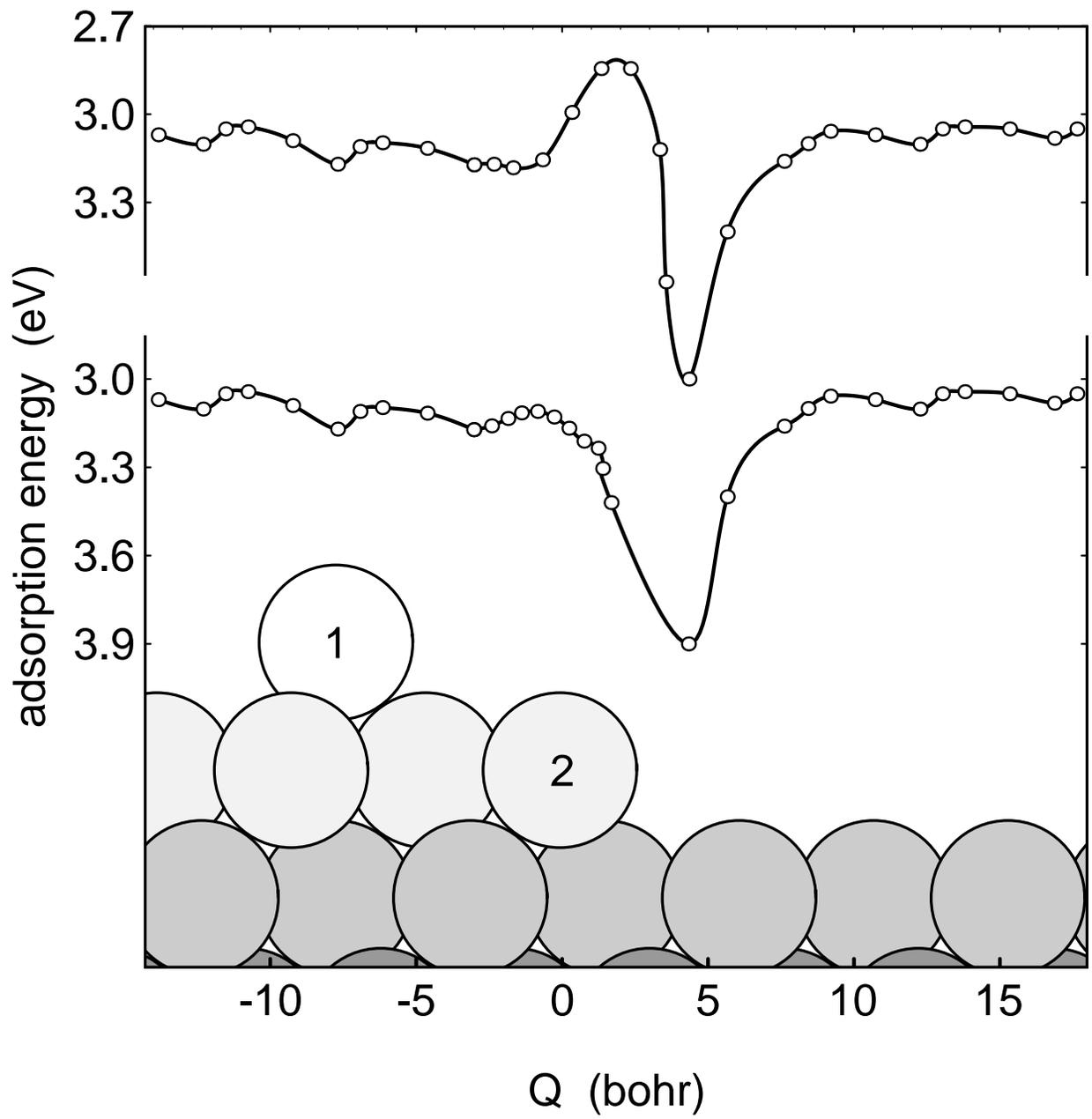
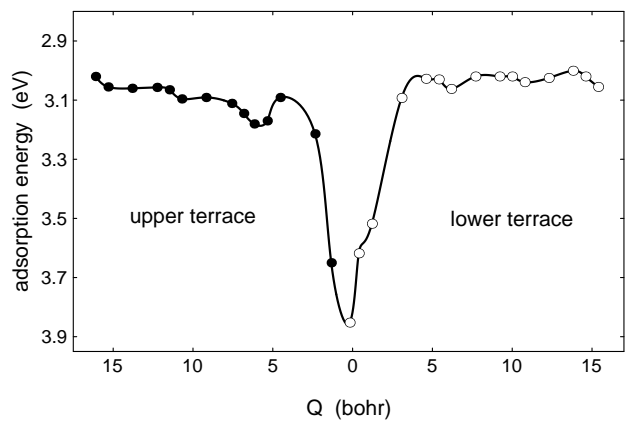
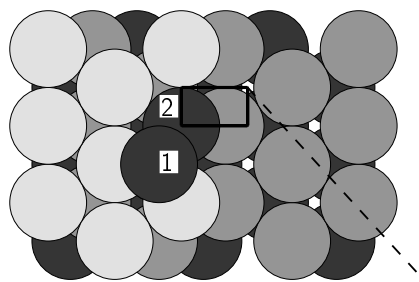
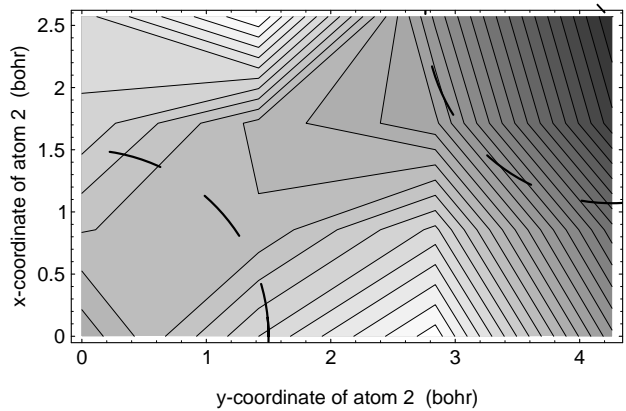
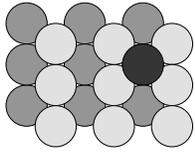
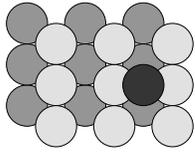
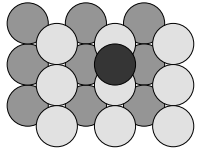
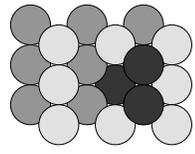
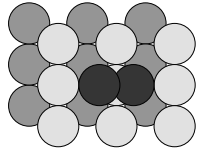
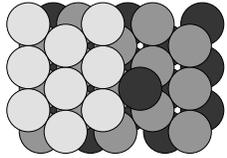
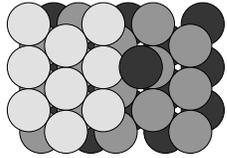
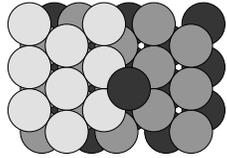
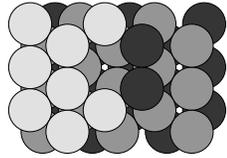
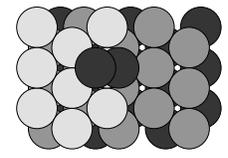
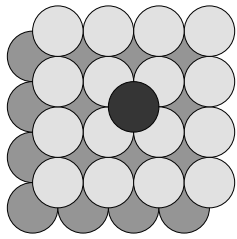
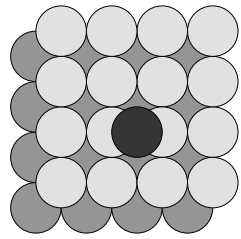
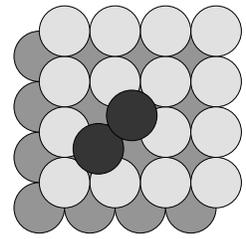
\end{document}